\documentclass[10pt]{iopart}
\usepackage{iopams}  
 \usepackage{amsmath}
\makeatletter
\@namedef{ver@amsmath.sty}{}
\makeatother
\usepackage{amsfonts}
\usepackage{amssymb}
\usepackage{color}
\usepackage{slashed}
\usepackage{enumerate}
\usepackage{graphicx}
\usepackage{bm}
\usepackage{braket}
\usepackage{epstopdf}

\usepackage[colorlinks=true,
            linkcolor=red,
            urlcolor=gray,
            citecolor=blue]{hyperref}
\begin{document}

\title[]{Study of vacuum behavior for inert models with discrete $Z_{2}$-like and abelian $U(1)$ symmetries}

\author{Andr\'es Castillo$^{1}$, Rodolfo A. Diaz$^{2}$, John Morales$^{3}$, Carlos G. Tarazona$^{4}$}
\ead{$^{1}$afcastillor@unal.edu.co,  $^{2}$radiazs@unal.edu.co, $^{3}$jmoralesa@unal.edu.co, $^{4}$caragomezt@unal.edu.co} 
\address{Departamento de F\'isica, Universidad Nacional de Colombia, Sede Bogot\'a}

\vspace{10pt}

\begin{abstract}
We study the vacuum behavior at one loop level in extended Higgs sectors with two doublets (2HDM), where $U(1)$ and $Z_{2}$ symmetries are considered to protect the $CP$ symmetry in the Higgs potential and to avoid Flavor Changing Neutral Currents at tree level in the Yukawa sector. In the Inert Higgs Model case, a detailed comparison is made between both models by using the energy evolution of couplings, which should satisfy energy scale dependent relations deduced for minima and stationary points of the Higgs potential at tree level. Besides, perturbative unitarity constraints at tree level are considered to generate the allowed parameter space compatible with perturbativity (absence of Landau poles). Our studies illustrate exclusion regions for Higgs masses and other combinations of couplings in the scalar sector, in particular for splittings of mass square for neutral scalars $A^{0}$ and $H^{0}$, as well as the difference between the sum of these and the charged Higgs mass square. From the vacuum stability for inert-2HDM at the tree and one loop levels, analyses lead us to find out new hierarchical structures for scalar masses.  To complete vacuum studies on the Inert model, and based on reparameterization invariance of the Higgs potential, we compute original discriminants that allow ensuring the presence of a global electroweak minimum at tree level. Moreover, the behavior in high energy scales drives out analyzing criticality phenomena for the additional parameters of extended Higgs sectors. Finally, and using the consistency with the electroweak precision analyses of oblique parameters, we describe several implications from different regimes of the inert model on charged and pseudoscalar Higgs searches.
\end{abstract}

\section{Introduction}

Scalar signal compatible with a Higgs hypothesis and favored by the
experimental data in CMS and ATLAS leads to a mass close to $125$ GeV \cite%
{CMSATLAS}-\cite{PDG}. This mass region has been studied comprehensively from vacuum analysis at next to leading order (NLO) \cite{NLO}-\cite{NLO8} and in the most contemporary analysis at next to next to leading order (NNLO) \cite{Degrassi}. The first approach relies on two loop renormalization group equations and one loop threshold corrections at the
electroweak scale improved with two loop terms from pure QCD corrections. On
the other hand, the NNLO incorporates higher order corrections in the strong, top Yukawa and Higgs quartic couplings; considering mainly full three loop beta functions for all SM gauge couplings and the leading terms three loop beta functions in the RG evolution. Moreover, NNLO terms have
an important piece of the vacuum stability analysis that comes from two-loop corrections to quartic coupling at the weak scale due to QCD and top Yukawa interactions, because such couplings are sizable at low energy scales. With these computations, absolute stability of the Higgs potential is excluded at $98\%$ C.L. for $m_{h}<126$ GeV while quartic coupling at the Planck scale is close to zero, which is associated with critical phenomena \cite{Degrassi}. Indeed, in the current mass region for Higgs and top quark, there is a significant preference for \emph{metastability} of the SM
potential \cite{MessinaI}. This situation takes place when the true minimum of the scalar potential is deeper than the standard electroweak minimum, but the latter has
a lifetime that is larger than the age of the universe \cite{Salvio,MessinaII}.

The occurrence of criticality (metastability-stability boundary) for couplings at higher energy scales could be a consequence of symmetry, a fine tuning or a dynamical effect among new parameters from an Extended Higgs Sector of the Inert Two Higgs Doublet Model (IHDM), for instance. The last one is the primary motivation for our work: try to
understand some limits of this criticality phenomena through of extended models respecting the minimality principle -wherein the typical scales for SM and the Inert Two Higgs Doublet Model are the same-. Moreover, those analyses would be involved in the threshold corrections in the study of phenomenology in other
models beyond SM sharing a similar Higgs spectrum (with the decoupling other particle states).

{Criticality in our case reflects as the boundary separating the stability and instability behaviors in the effective potential for the extended Higgs sector. By plotting different combinations of new quartic couplings of extended parameter space, we analyze the behavior of criticality with energy scales. Since these couplings also depend on measurements for Higgs boson and top quark masses, the improvement for the precision level of these parameters leads to describing most accurately mechanisms and principles behind of phase diagrams.}

Before discussing particular inert models, we point out that general 2HDMs provide a general effective theory framework for extensions of the electroweak symmetry breaking sector, supersymmetric among others. The 2HDMs includes two complex doublets with identical quantum numbers. By counting the degrees of freedom introduced by the new doublet,  in the 2HDMs there are
eight real fields: three must become the longitudinal components of the $%
W^{\pm }$ and $Z^{0}$ bosons after the spontaneous symmetry breaking. Five physical Higgs scalars will remain: a charged scalar $%
H^{\pm }$ and three neutral scalars $h^{0},H^{0}$ and another neutral
pseudoscalar $A^{0}$. Some of the main motivations to introduce an extended Higgs
sector with two doublets are the sources for either explicit or spontaneous CP violation. Other phenomenological aspects and several experimental searches for the general 2HDMs in the light of different LHC results have been treated in very illustrative papers \cite{Report Sher}-\cite{Djouadi}.

Moreover, 2HDMs contain parameter spaces compatible with Baryon Asymmetry of the Universe (BAU) \cite{BAU}. In this direction, the
non-compatibility of SM dynamics with Sakharov conditions, notably the absence of a strong first order phase transitions, is a significant failure of the minimal model with one doublet \cite{BAUI}. Mechanisms to take into account BAU through Sakharov conditions realization require a complete study of vacuum structures of extended models, motivating all possible studies about stability beyond SM.

Another major consequence with new Higgs doublets is the possibility of flavor-changing neutral currents (FCNC). It is well known that FCNCs are highly constrained concerning charged current processes like mesons oscillations, so it would be desirable to \textquotedblleft naturally\textquotedblright suppress them in these models. If all fermions with the same quantum numbers are coupled to the same scalar doublet, then FCNCs will be absent. A necessary and sufficient condition leading to an absence of FCNC at tree level is that all fermions of a given charge and helicity transform according to the same irreducible representation of $SU(2)$ group, corresponding to the same eigenvalue of the third component of isospin operator. Thus exist a basis in which fermions receive their contributions in the
mass matrix from a single source \cite{Weinberg,Paschos}. To achieve this effect in the quark sector of 2HDM,
we can see two possibilities: all quarks couple to just one doublet (here it
has been chosen to be $\Phi _{1}$)\footnote{
Due to the choice of an inert vacuum in $\langle \Phi_{2}\rangle_{0}=0$, we take the fermion sector-couplings coming from of the first doublet.} or the up type
right-handed quarks couple to one doublet (e.g. $\Phi _{1}$) and the down
type right-handed quarks couple to the other ($\Phi _{2}$). The former model is
called the  type I 2HDM, meanwhile the last model is known as the type
II-2HDM. When these structures are extended to the leptonic sector, it is assumed
the charged leptons couple to the same Higgs doublet as the $Q=-1/3$ quarks,
although this condition is not unique. Indeed there are at least
other two possibilities to build up models with natural flavor conservation.
In the lepton specific model, for instance, the RH quarks couple to $\Phi _{2}$ and the RH
leptons couple to $\Phi _{1}.$ In the flipped model, the $Q=2/3$ right-handed quarks and charged leptons couple to the same doublet (say $\Phi
_{1} $), and the $Q=-1/3$ right-handed quarks are coupled to $\Phi _{2}$ 
\cite{Report Sher}. In this work we are only focused on the traditional type
I 2HDM and the corresponding extension to lepton sector \footnote{%
The Yukawa Lagrangians type I and II can also be generated from a continuous global symmetry. The set of transformations 
\begin{eqnarray*}
\Phi _{1} &\rightarrow &e^{i\varphi }\Phi _{1}\text{ and }\Phi
_{2}\rightarrow -\Phi _{2} \\
D_{jR} &\rightarrow &e^{-i\omega }D_{jR}\text{ and }U_{jR}\rightarrow
e^{-i\varphi }U_{jR}
\end{eqnarray*}%
with $\omega =\varphi ,\pi /2,$ yield models type I and type II respectively. Here $D_{R}$ is refers to the three down type weak
isospin quark singlets and $U_{R}$ is assigned to the three up type weak
isospin quark singlets.}, because the symmetries considered
are achieved exactly either in the Higgs potential or the Yukawa sector.

Since the Higgs mass in SM is becoming constrained even more
through precision tests performed in LHC, one might ask how the remaining scalar free parameters in the 2HDM are limited by the general vacuum behavior \cite{Sher-Nie}. Indeed, there are additional
quartic-couplings and thus more directions in field space where instabilities in the Higgs potential could appear. Moreover, self-couplings evolution could lead to Landau Poles and non-perturbative regimes.  Consequently, limits over model parameters or splittings (mass differences) depend on mass eigenstates
structure in 2HDM and therefore on symmetries in the Higgs potential; converting those splittings into a crucial tool to make phenomenological analyses of compatibility.
Particularly, one first scenario of extended Higgs sectors with outstanding vacuum structures and with phenomenological consequences is the inert Higgs doublet model. In this work, we examine these constraints in the IHDM with $U\left( 1\right), Z_{2} $ global symmetries, under a choice of
vacuum where $\Phi _{2}$ doublet has a VEV equal to zero. This parametrization is made in models with natural flavor conservation\footnote{In 2HDM type II, selection of inert vacuum prevents to down-type fermions to
acquire mass. By virtue we are also interested in to study effects of heaviest down-type fermions in RGEs, we would study only the 2HDM type I.}. To restrict charged and pseudoscalar masses (or splitting between them), we can identify a Higgs state $h^{0}$ with the current signal for Higgs boson and study the remaining directions in the parameter space. Our work is a  complementary study to the constraints of the parameter space of 2HDM of different mass splittings (with new effects to get an EW-global minimum at tree level), which has been examined before \cite{VEarticles, VEarticles2} in non-inert scenarios. The most current studies in this direction for inert and non-inert models were carried out in \cite{FerreiraMet}-\cite{Inertall}.

Additional features of benchmarks associated with the inert-Higgs doublet
model have been introduced to set a heavier Higgs boson $H^{0}$ of mass running between 400 and 600 GeV. That would lift the divergence of the Higgs mass
radiative corrections beyond the TeV scale, where new physics is considered to provide a compelling naturalness in theory and to make the Higgs quartic coupling perturbative \cite{Barbieri}. Since the perturbative unitarity bounds for $m_{H^{0}},$ $m_{A^{0}}$ and $m_{H^{\pm }}$ are near to $700$ GeV for models compatible with an inert 2HDM \cite{Kanemura}-\cite{Kladiva}, the interval to interpret naturalness problem is also in agreement with this unitarity tree-level bound. On the other hand, constraints from the precision electroweak data embodied in the oblique parameters encourage using a second inert Higgs doublet $\Phi_{2}$. Despite $\Phi _{2}$ has scalar interactions just as in the ordinary 2HDM, it does not acquire a vacuum expectation value (its
minimum is at $(0,0)$), nor it has any other couplings to fermions and gauge bosons. The
inert-2HDM, therefore, belongs to the class of Type-I 2HDMs introduced above; in our particular case $\Phi_{1}$ saturates couplings with fermions and gauge bosons. This fact ensures an alignment regime where $h^{0}$ behaves as SM-like Higgs boson \cite{Carena}.

Indeed, the inert doublet can have an odd parity under an unbroken intrinsic $Z_{2}$ symmetry, while all the SM fields have even $Z_{2}$ parity. This transformation translates the lightest inert scalar  ($H^{0}$ or $A^{0}$) into in a stable and a viable dark matter candidate \cite{STIHDM, Tytgat}. For this framework, vacuum constraints at tree level, relations to avoid minima with charge violation and phenomenology in LHC have been studied in \cite{KrawczykII} obtaining a particular organization of the scalar spectrum. Despite the fact that the archetype for an inert model is the $Z_{2}$ invariant theory, the present paper is also focused on the impact of a Higgs potential with a $U(1)-$symmetry, which
also has significant phenomenological consequences for dark matter searches \cite{DM}. Furthermore, in this direction, we find out distinctions and ways to discriminate both models. Although splittings are not commonly constrained, similar analysis from vacuum stability, general alignment regime and unitarity in the context of non-inert 2HDM with softly broken symmetries for independent parameters can be found in \cite{Dipankar}. A crucial point from these analyses is the relevance of softly breaking parameters in discriminating of stable or unstable zones along parameter space compatible with a scalar alignment regime.  

Our paper is organized as follows: In section \ref{sec:Inert}, we discuss  particular cases of $Z_{2}$ and $U\left(
1\right)$ global symmetries of the Inert Two Higgs Doublet Model. Additionally, we after discussing positivity constraints as well as conditions for the presence of a global minimum in the Higgs potential at tree level. Mass eigenstates and splittings among scalars in the inert 2HDM will be given in the same section. In section \ref{sec:Unitarity}, we describe perturbative unitarity constraints to the scalar sector for both models ($Z_{2}$ and $U(1)$). In section \ref{sec:VBC}, contours and the corresponding analyses of couplings are considered in different energy scales, from Electroweak up to GUT (Grand Unification Theories) and
Planck scales, for type I Yukawa Lagrangian. At the same time in those studies, we find compatibility with perturbative unitarity behavior for scalar couplings. Tree level regions compatible to get one global electroweak minimum are described in section \ref{sec:metastabilityanalyses}. According to the restrictions obtained, in section \ref{sec:ST}, oblique electroweak parameters are computed to establish the compatibility between vacuum behavior predictions and
phenomenological observables. Finally, in the conclusions and remarks, we
discuss the influence of our treatment in the interpretation of vacuum analysis and the compatibility with these EW precision tests.

\section{Inert Two Higgs Doublet Model (IHDM)}
\label{sec:Inert}

Preserving the SM content of fermionic and bosons fields, the Inert Two Higgs Doublet Model contains additionally a doublet $\Phi_{2}$ with a VEV equal to zero. The model has a general $Z_{2}$-invariance, under which $\Phi_{2}$ transforms odd, and the remaining fields change even. At tree level, $\Phi_{2}$ does not couple with fermions. The physical parametrization of the Higgs doublets is
\begin{align}
 \Phi_{1}=\begin{pmatrix}
           G^{+}\\
           \frac{1}{\sqrt{2}}\left(v+h^{0}+iG^{0}\right)  
          \end{pmatrix}\text{ and }\Phi_{2}=\begin{pmatrix}
           H^{+}\\
           \frac{1}{\sqrt{2}}\left(H^{0}+iA^{0}\right)
          \end{pmatrix}, \label{Parametrization}
\end{align}

featuring five Higgs bosons ($h^{0},H^{0},A^{0},H^{\pm}$) and three Goldstone bosons ($G^{0},G^{\pm}$). The vacuum expectation value for the first doublet is located in $\langle \Phi_{1}\rangle_{0}= v=246$ GeV. Fields $h^{0}$ and $H^{0}$ are defined as scalars transforming to $CP$ symmetry in a even way, meanwhile $A^{0}$ is a pseudoscalar field changing odd under $CP$ symmetry. Finally, fields $H^{\pm}$ are the charged Higgs bosons. 

The scalar field $h^{0}$ emulates SM Higgs boson in mass and couplings with fermionic and gauge bosonic fields, trivially satisfying an \emph{alignment regime} in the scalar sector\footnote{In the general alignment regimen, the remaining scalars can be located in any energy scale fulfilling the electroweak oblique parameters \cite{Carena}}. The Higgs potential in this context takes the following form:

\begin{align}
V_{H}& =m_{11}^{2}\Phi _{1}^{\dagger }\Phi _{1}+m_{22}^{2}\Phi _{2}^{\dagger
}\Phi _{2}+\frac{1}{2}\lambda _{1}\left( \Phi _{1}^{\dagger }\Phi
_{1}\right) ^{2}+\frac{1}{2}\lambda _{2}\left( \Phi _{2}^{\dagger }\Phi
_{2}\right) ^{2}\notag\\
&+\lambda _{3}\left( \Phi _{1}^{\dagger }\Phi _{1}\right)
\left( \Phi _{2}^{\dagger }\Phi _{2}\right) +\lambda _{4}\left( \Phi
_{1}^{\dagger }\Phi _{2}\right) \left( \Phi _{2}^{\dagger }\Phi _{1}\right) 
 + \frac{1}{2}\lambda _{5}\left[\left( \Phi _{1}^{\dagger }\Phi
_{2}\right) ^{2}+\left( \Phi _{2}^{\dagger }\Phi
_{1}\right) ^{2}\right].  \label{PotV1}
\end{align}

We have considered a $CP$ conserving Higgs potential by taking all couplings in (\ref{PotV1}) as real quantities. Under an Abelian theory, a global $U(1)$-symmetry excludes $\lambda_{5}$ coupling in $V_{H}$. If we choose an inert second doublet, i.e. $\langle \Phi _{2}\rangle _{0}=0,$
Higgs masses acquire the following structure: %

\begin{eqnarray}
m_{h^{0}}^{2} &=&\lambda _{1}v^{2},  \label{Ei1} \\
m_{H^{0}}^{2} &=&m_{22}^{2}+\frac{1}{2}\lambda _{3}v^{2}+\frac{1}{2}\left(
\lambda _{4}-\lambda _{5}\right) v^{2}+\lambda
_{5}v^{2}=m_{A^{0}}^{2}+\lambda _{5}v^{2},  \label{Ei2} \\
m_{A^{0}}^{2} &=&m_{22}^{2}+\frac{1}{2}\lambda _{3}v^{2}+\frac{1}{2}\left(
\lambda _{4}-\lambda _{5}\right) v^{2}=m_{H^{\pm }}^{2}+\frac{1}{2}\left(
\lambda _{4}-\lambda _{5}\right) v^{2},  \label{Ei3} \\
m_{H^{\pm }}^{2} &=&m_{22}^{2}+\frac{1}{2}\lambda _{3}v^{2}.  \label{Ei4}
\end{eqnarray}

{From this settlement of equations, we realize that the mass eigenstates are independent of $\lambda _{2}.$ This fact prevents to constraint $\lambda _{2}$ with phenomenology for scalar boson $h^{0}$ directly because production or decay rates with $\lambda_{2}$ depend on new physics Higgs bosons $H^{0}, A^{0}$ and $H^{\pm}$. This fact motivates to vacuum stability and perturbativity analyses since these approaches are meaningful ways to give feasible values for this particular coupling}. Besides, $\lambda _{5}$ coupling prevents mass degeneracy between $H^{0}$ and $%
A^{0}$ scalars. This regime of degeneracy will be present in a Higgs potential with $%
U\left( 1\right) $ symmetry \footnote{Moreover, extending the $Z_2$ to a global $SU(2)$ acting on $\Phi_{2}$ makes both $\lambda_{4}=\lambda_{5}=0$ and forces
all three inert scalars degenerate. This fact motivates a \emph{compressed}-IHDM where all inert scalars have almost degenerated masses and where $SU(2)$ global symmetry is an approximated symmetry of the Higgs potential \cite{DelaPuente}.}. In the last scenario, a remarkable fact observed is the non-appearance of an axion with $m_{A^{0}}=0$ (emerging when a continuous global symmetry becomes spontaneously broken), which is due to the choice of an inert doublet makes that the $U(1)$-global symmetry remains unbroken.

Because of the relation of $\lambda_{5}$ with the scalar masses, it is possible to define
the splitting among masses of pseudoscalar and the heaviest neutral Higgs by %

\begin{equation}
\lambda _{5}=\frac{m_{H^{0}}^{2}-m_{A^{0}}^{2}}{v^{2}}\equiv \Delta
S_{0}^{2}.
\end{equation}

For $\lambda_{4}$, relation with the scalar masses induces a splitting between neutral and
charged scalars:

\begin{equation}
\lambda _{4}=\frac{m_{H^{0}}^{2}+m_{A^{0}}^{2}-2m_{H^{\pm }}^{2}}{v^{2}}%
\equiv \Delta S_{1}^{2}.
\end{equation}

It is also convenient to define the difference between charged Higgs mass and $m_{22}^{2}$ parameter

\begin{align}
 \lambda_{3}=\frac{2\left(m_{H^{\pm}}^{2}-m_{22}^{2}\right)}{v^{2}}\equiv\Delta S_{2}^{2}.
\end{align}

\subsection{Vacuum Stability Behavior}
\label{sec:TL}
To ensure a bounded from below Higgs potential, it is necessary the exigence that $V_{H}$ in Eq. (\ref{PotV1}) must always be positive for large field values along all possible directions of the $(\Phi_{1},\Phi_{2})$ space. At tree level, this is translated into the following inequalities \cite{Sher-Nie,Ivanov,Ivanov4}

\begin{align}
\lambda_{1}+\lambda_{2}>\vert \lambda_{1}-\lambda_{2}\vert, \label{stab1}
\end{align}

which is equivalent to $\lambda_{1}>0$ and $\lambda_{2}>0$ in the individual directions of $\Phi_{1}$ and $\Phi_{2}$ directions. In the plane $\Phi_{1}-\Phi_{2}$, the positivity conditions are

\begin{align}
 \lambda_{3}&>-\sqrt{\lambda_{1}\lambda_{2}},\\
 \lambda _{4}+\lambda _{3}+\lambda _{5} &>-\sqrt{\lambda _{1}\lambda _{2}},
\\
\lambda _{4}+\lambda _{3}-\lambda _{5} &>-\sqrt{\lambda _{1}\lambda _{2}}.\label{stab3}
\end{align}

{These inequalities (\ref{stab1})-(\ref{stab3}) ensure absolute stability for the electroweak vacuum by defining a bounded from below Higgs potential. However, possible metastable scenarios arise when a second inert-like extremum is specified by a VEV $v_{2}$ non-zero and $v_{1}=0$}. In this stationary point, the $Z_2$ symmetry of the Higgs potential is conserved by this state; however, the $Z_2$ symmetry of the Lagrangian become spontaneously violated. In this framework, fermions are massless since they uniquely couple to $\Phi_{1}$. Therefore, this non-physical behavior must be excluded from a plausible parameter space when this extremum point become one global minimum of the theory. Two necessary conditions for the simultaneous existence of both minima is that i) $m_{11}^{2}<0$ and $m_{22}^{2}<0$ or ii) $\lambda_{3}+\lambda_{4}+\lambda_{5}>0$ \cite{FerreiraMet}.

One condition to ensure that the inert vacuum would be the global minimum of the Higgs potential is \cite{Ivanov,Kanishev}

\begin{align}
 \frac{m_{11}^{2}}{\sqrt{\lambda _{1}}}<\frac{m_{22}^{2}}{\sqrt{\lambda _{2}}}.
\end{align}

To determine if the EW-minimum (inert) is a global one, we calculate a new set of inequalities relating to quartic and bilinear couplings with critical points in the Higgs potential.
The new discriminants, encouraging a global minimum in the Higgs potential, are computed for IHDM from the respective Hessian in the gauge orbit field using the general reparameterization group $SO(1,3)^{+}$ evaluated in the inert-stationary point\footnote{Our computations are based on new methods for searching stationary points in 2HDMs, which are defined systematically in \cite{Ivanovmet2015}}:

\begin{align}
 -\sqrt{\lambda_{1}\lambda_{2}}<\frac{2m_{22}^{2}}{v^{2}}<\sqrt{\lambda_{1}\lambda_{2}},\label{Meta1}\\
 -(\lambda_{3}+\lambda_{4}+\lambda_{5})<\frac{2m_{22}^{2}}{v^{2}}<\sqrt{\lambda_{1}\lambda_{2}},\\
 -(\lambda_{3}+\lambda_{4}-\lambda_{5})<\frac{2m_{22}^{2}}{v^{2}}<\sqrt{\lambda_{1}\lambda_{2}}.\label{Meta3}
\end{align}

We focus only on the implications that new discriminants have over parameters at tree level. Possible studies might be done in the future by exploring the consequences at one loop level since many phenomena over nature of minima seem to show intriguing effects of the effective Higgs potential at NLO\cite{FerreiraMet}.

Despite in 2HDMs at tree level two minima that break different symmetries cannot coexist, a global minimum with charge violation can appear if quartic couplings satisfy \cite{Kanishev2}-\cite{Kanishev3}

\begin{align}
\lambda_{4}-\lambda_{5}>0 \text{ and } \lambda_{5}+\lambda_{4}>0 \text{ and } \lambda_{3}-\sqrt{\lambda_{1}\lambda_{2}}>0.
\end{align}

Mass eigenstates and conditions to avoid charge violation vacua lead to study possible sequences for scalar masses. For instance, from $\lambda _{4}+\lambda _{5}<0,$ we can infer the hierarchy $m_{H^{\pm }}>m_{H^{0}}$ for scalar masses, which is also inherited by a $%
U\left( 1\right) $ Higgs potential with the additional consequence of degeneracy
between $H^{0}$ and $A^{0}.$ By contrast, $\lambda_{4}-\lambda_{5}>0$ implies $m_{A^{0}}>m_{H^{\pm }}$ for the $Z_{2}$ invariant model. With these constraints in mind, we shall study their compatibility level with stability and unitarity bounds.

\section{Unitarity constraints}
\label{sec:Unitarity}

Unitarity constraints at tree level arise using the optical theorem in the $S$-matrix description with generalized partial waves for scalar scattering processes.  The traditional way to implement it in gauge theories is demanding that the model has only weakly interacting degrees of freedom at high energy limit. In these weakly coupled theories, higher order contributions to $S$-matrix become smaller compared to the leading order. It is then possible to require for $S$-matrix to be unitary from the tree level of the theory.

We concentrate on scalar processes coming from Higgs potentials
with $U(1)$ and $Z_{2}$ symmetries. In a generic basis of the Higgs
potential Eq. (\ref{PotV1}), processes labeling total isospin $\sigma$ and
hypercharge $Y$ lead to construct the following
transition matrices for the particular case of a $Z_{2}$ invariant Higgs potential \cite{Ivanov3}:

\begin{subequations}
\label{eigenvalues}
\begin{align}
8\pi \tilde{S}_{Y=2,\sigma =1}& =%
\begin{pmatrix}
\lambda _{1} & \lambda _{5} & 0 \\ 
\lambda _{5} & \lambda _{2} & 0 \\ 
0 & 0 & \lambda _{3}+\lambda _{4}%
\end{pmatrix}%
, \\
8\pi \tilde{S}_{Y=2,\sigma =0}& =\lambda _{3}-\lambda _{4},  \label{b} \\
8\pi \tilde{S}_{Y=0,\sigma =1}& =%
\begin{pmatrix}
\lambda _{1} & \lambda _{4} & 0 & 0 \\ 
\lambda _{4} & \lambda _{2} & 0 & 0 \\ 
0 & 0 & \lambda _{3} & 0 \\ 
0 & 0 & 0 & \lambda _{3}%
\end{pmatrix}%
, \\
8\pi \tilde{S}_{Y=0,\sigma =0}& =%
\begin{pmatrix}
3\lambda _{1} & 2\lambda _{3}+\lambda _{4} & 0 & 0 \\ 
2\lambda _{3}+\lambda _{4} & 3\lambda _{2} & 0 & 0 \\ 
0 & 0 & \lambda _{3}+2\lambda _{4} & 3\lambda _{5} \\ 
0 & 0 & 3\lambda _{5} & \lambda _{3}+2\lambda _{4}%
\end{pmatrix}%
.  \label{matricesdedispersion}
\end{align}
\end{subequations}

In each matrix, left sides contain matrix elements for $Z_{2}-even$ states,
meanwhile in the lower right part they belong to $Z_{2}-odd$ states. Unitarity bounds over partial waves $|\mathcal{M}|<1$ are translated into eigenvalues $\mathbf{\Lambda}$ for $%
\tilde{S}$ matrices, hence the new condition is

\begin{equation}
|\boldsymbol{\Lambda}|<\frac{1}{8\pi \xi},
\end{equation}

with $\xi$ an indistinguishability factor. As was discussed above, this upper bound corresponds to equal the $\mathcal{M}$-matrix with the tree-level elements by disregarding higher order corrections. For the Higgs potential (\ref{PotV1}), the matrices in (\ref{eigenvalues}) are block-diagonal facilitating the computation of the eigenvalues $\boldsymbol{\Lambda} _{Y,\sigma \pm
}^{Z_{2}}$:

\begin{subequations}
\begin{align}
\boldsymbol{\Lambda} _{2,1\pm }^{even}& =\frac{1}{2}\left( \lambda _{1}+\lambda _{2}\pm 
\sqrt{(\lambda _{1}-\lambda _{2})^{2}+4\lambda _{5}^{2}}\right) ,\hspace{%
0.4cm}\boldsymbol{\Lambda}_{21}^{odd}=\lambda _{3}+\lambda _{4}.  \label{eigen1} \\
\boldsymbol{\Lambda}_{2,0\pm }^{even}& =\lambda _{3}-\lambda _{4}. \\
\boldsymbol{\Lambda}_{0,1\pm }^{even}& =\frac{1}{2}\left( \lambda _{1}+\lambda _{2}\pm 
\sqrt{(\lambda _{1}-\lambda _{2})^{2}+4\lambda _{4}^{2}}\right) ,\hspace{%
0.4cm}\boldsymbol{\Lambda} _{01\pm }^{odd}=\lambda _{3}\pm \lambda _{5}. \\
\boldsymbol{\Lambda}_{0,0\pm }^{even}& =\frac{1}{2}\left[ 3(\lambda _{1}+\lambda
_{2})\pm \sqrt{9(\lambda _{1}-\lambda _{2})^{2}+4\left( 2\lambda
_{3}+\lambda _{4}\right) ^{2}}\right] ,\hspace{0.4cm}\boldsymbol{\Lambda}_{00\pm
}^{odd}=\lambda _{3}+2\lambda _{4}\pm 3\lambda _{5}.  \label{eigen4}
\end{align}
\end{subequations}

These constraints will be used to see the compatibility between vacuum predictions and perturbative unitarity, as well as relationships with
the possible presence of Landau poles in the parameter space.

\section{One loop level analysis}

\label{sec:VBC}

\begin{figure}[htp]
\centering
\includegraphics[scale=0.25]{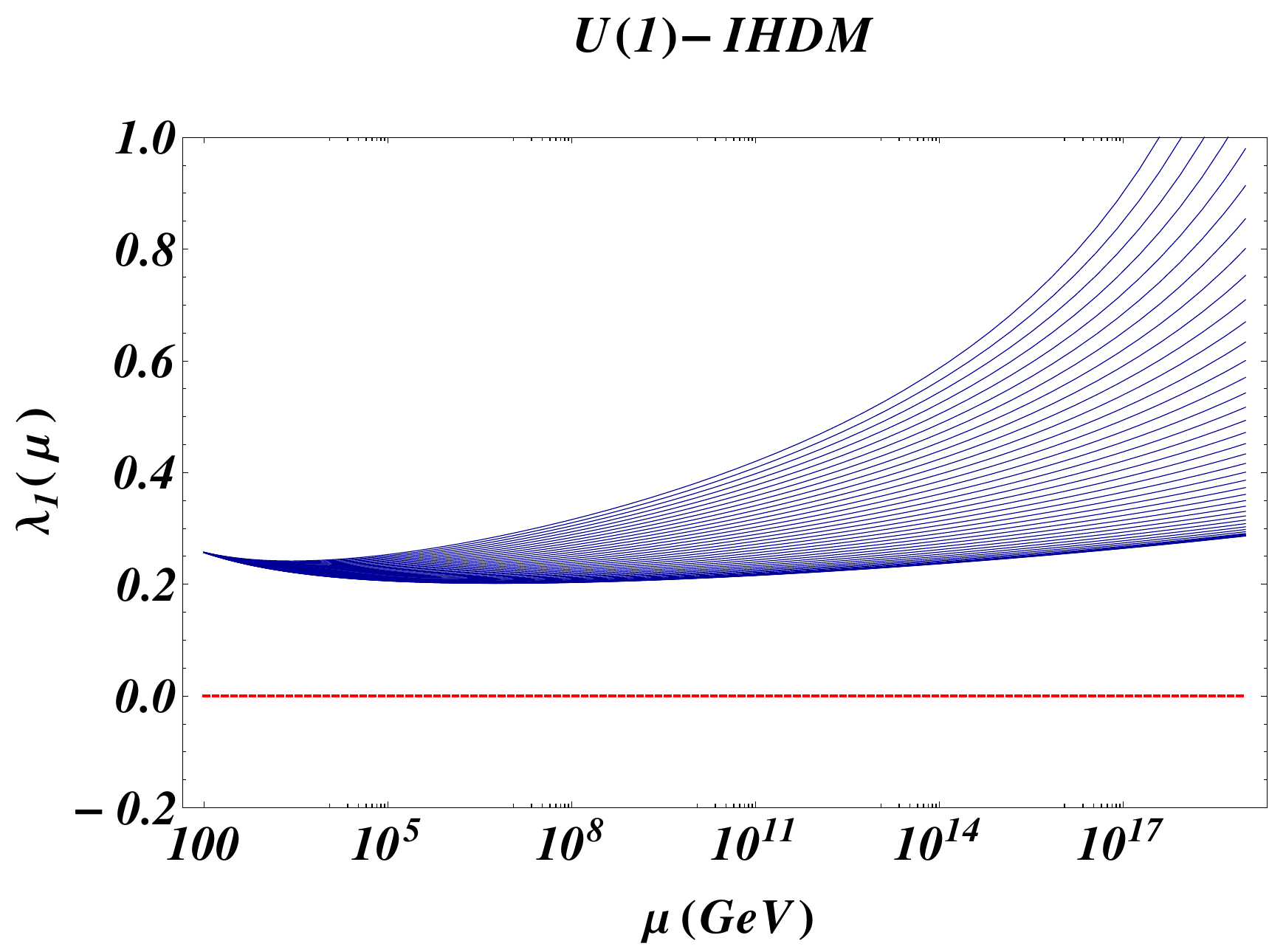}
\includegraphics[scale=0.242]{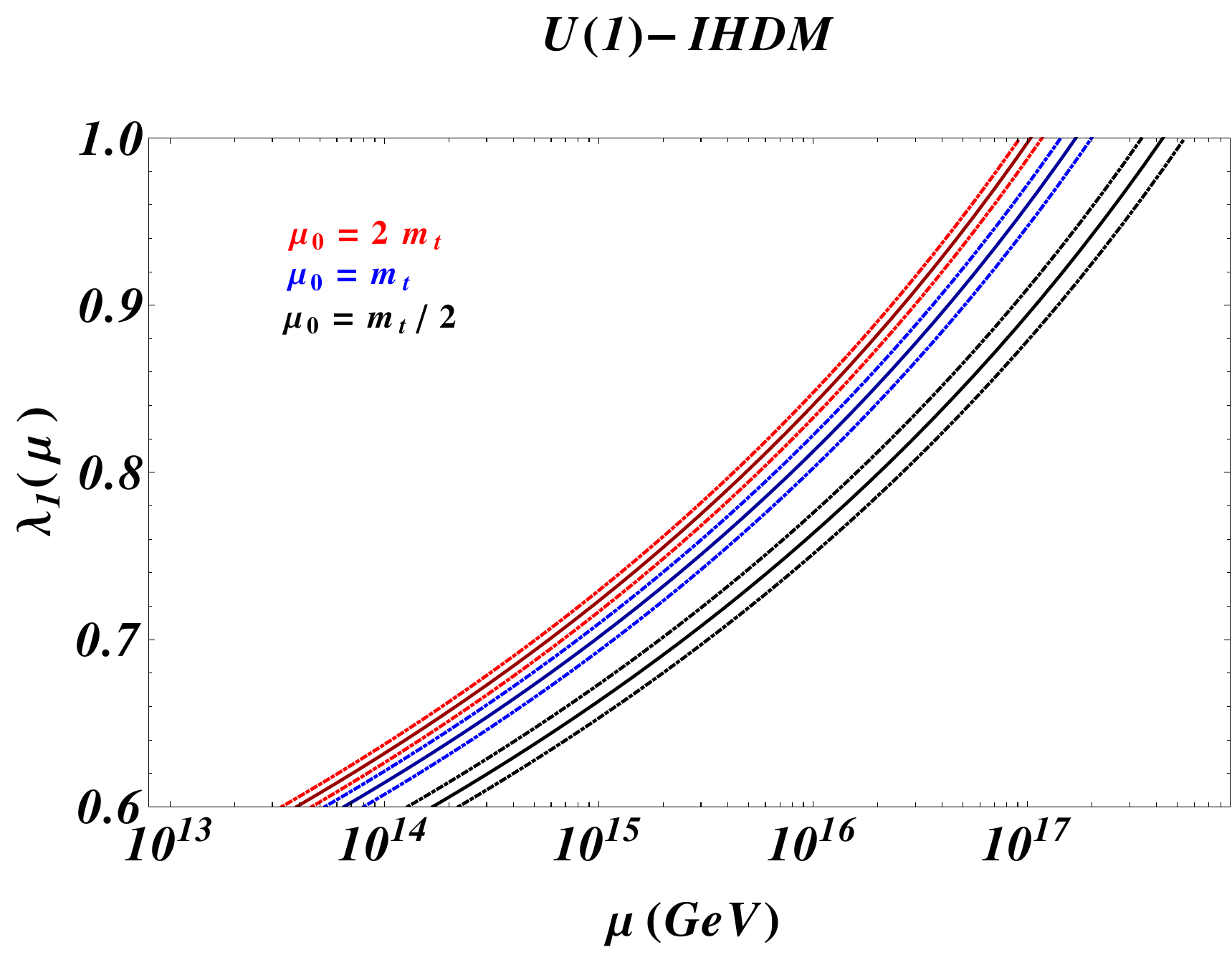} %
\vspace{0.6cm}
\caption{{(\textbf{Left}) Energy scale evolution for $\protect\lambda_{1}$ coupling with $m_{h^{0}}=125.04$ GeV and $
m_{t}=173.34$ GeV. Evolution of $\protect\lambda _{1}(\protect\mu )$ is made fixing the remaining initial conditions $\protect%
\lambda _{i}(\mu_{0}=m_{z} )$'s.  Initial gauge couplings have been taken at $m_{Z}$-scale. Here $-0.4\leq \protect\lambda _{4}(m_{Z})\leq 0.0$, $%
-0.2\leq \protect\lambda _{3}(m_{Z})\leq 0$ and $0.0\leq \protect\lambda %
_{2}(m_{Z})\leq 0.2$, and the assumption of $\lambda_{3}(m_{z})=\vert\lambda_{4}(m_{z})\vert/2$ and $\lambda_{3}(m_{z})=\lambda_{4}(m_{z})/2$. Each curve is varying in 0.02 units in those intervals. (\textbf{Right}) Maximum curve lying in such ranges that
specifies the top quark mass uncertainty \cite{Top} ($m_{t}=173.34 \pm 0.76$ GeV) leading to corrections for  $\lambda_{1}(\mu)$ ($\mu=10^{17}$ GeV) around $4\%$. Henceforth, numerical analysis are based on these central values of Higgs boson and top quark masses, likewise for gauge couplings in $m_{Z}$ scale \cite{PDG} and we have varied the matching condition $\mu_{0}=\{\frac{m_{t}}{2},m_{t},2m_{t}\}$. To assess
the uncertainty on the coupling, we have followed the prescription presented in \cite{DegrassiUncer}}}
\label{fig:EnergyEv}
\end{figure}

As a first proof of the influence of the scalar extended Higgs sector in the vacuum stability scenario, we consider the running coupling for $\lambda _{1}$ which could be compared with $\lambda _{SM}(\mu )$ through the appropriate limits of the theory. Indeed the numerical evaluation of Renormalization Group Equations (RGEs) for the 2HDM type I (they are depicted in \ref{ap:RGEsection}) allows computing the vacuum behavior and perturbative realization in field and parameter space for an $U(1)$-invariant model. This evolution can also be seen in a $Z_{2}$ invariant model with $\lambda_{5}(m_{Z})=0$. Figure %
\ref{fig:EnergyEv}- shows energy scale evolution for $\lambda _{1}(\mu )$
with different values of the remaining scalar couplings. $\lambda _{2}(\mu _{0})$ coupling is settled in such a way that its vacuum constraint satisfies, i.e. $%
\lambda _{2}(\mu _{0})>0$ (with $\mu_{0}=m_{Z}$ the initial scale). For the initial conditions taken there over
other couplings, the vacuum instabilities are suppressed in the $\Phi _{1}$
direction arise among $10^{3}-10^{19}$ GeV. 
Then, criticality presented in the SM at energies close to GUT and Planck scales for values of $\lambda _{1}(m_{Z})$ could be avoided in an extensive regime of the parameter space for an inert 2HDM type I.

\begin{figure}[htp]
\centering
\includegraphics[scale=0.18]{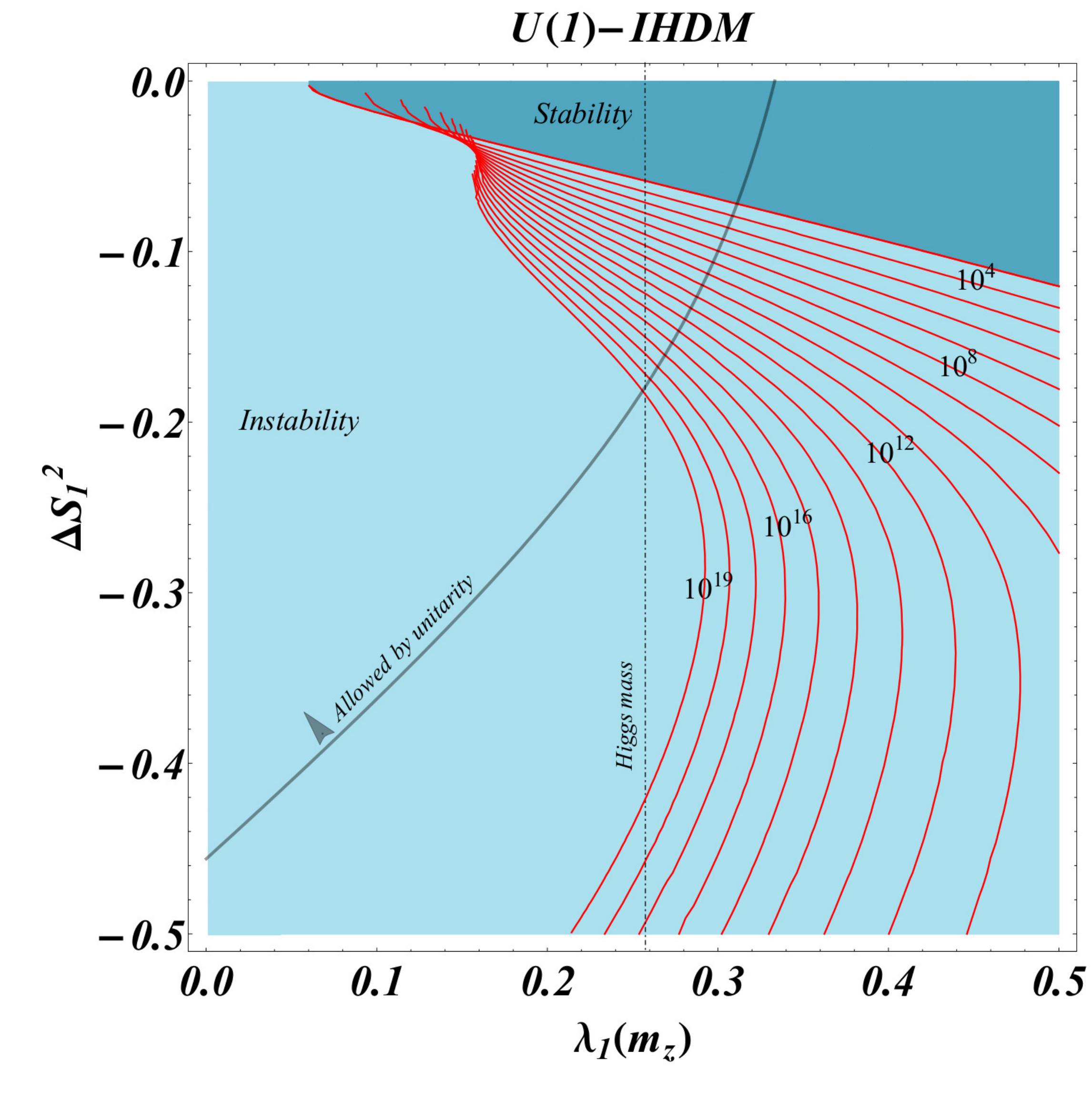} %
\includegraphics[scale=0.18]{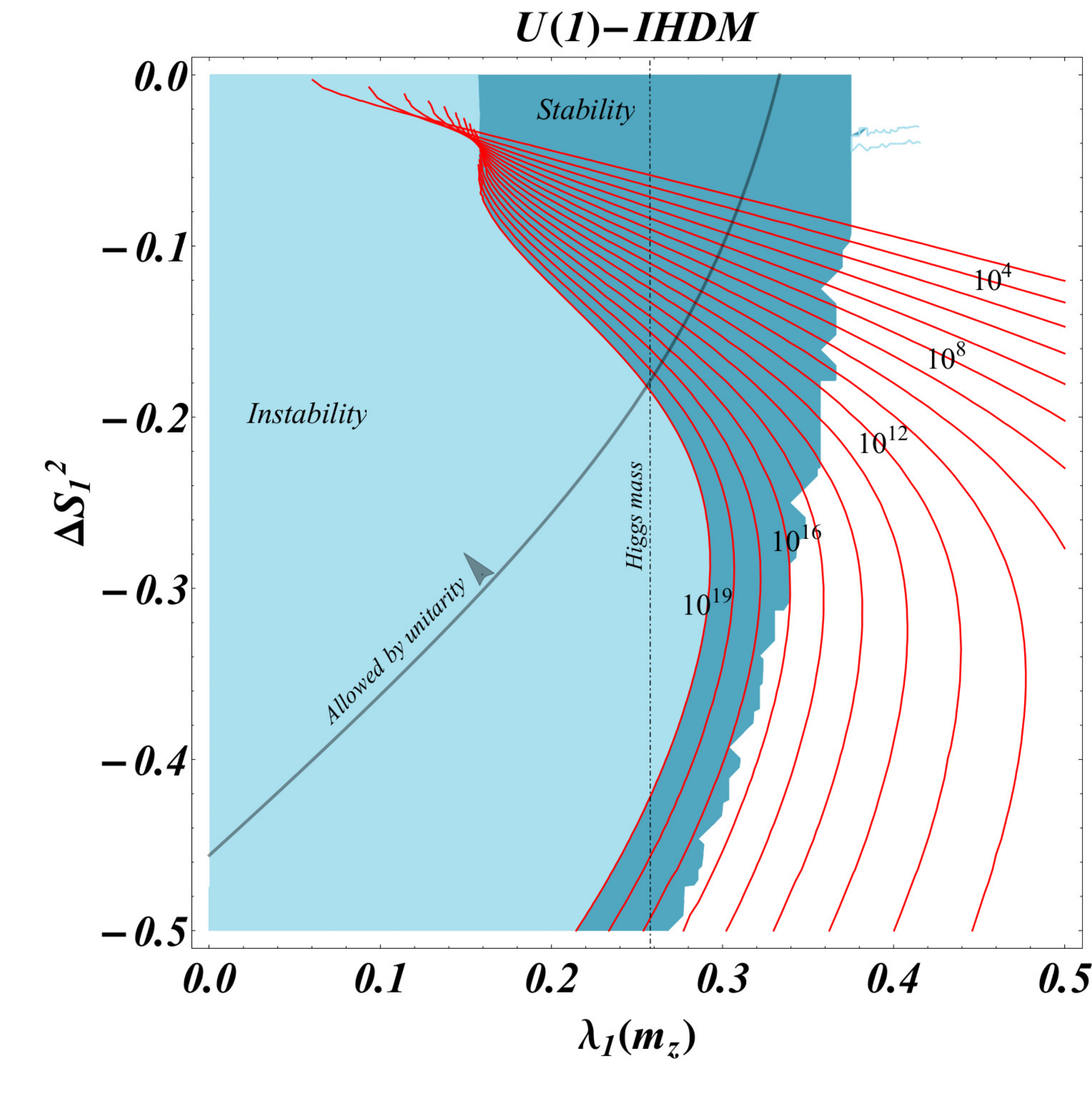} \vspace{0.6cm}
\caption{Phase diagrams with the evolution of contours from $\protect\mu %
=10^{3}$ GeV (\textbf{Background-Left}) up to $\protect\mu =10^{19}$ GeV (\textbf{Background-Right}) in the $\Delta S_{1}^{2}$ versus $\protect%
\lambda _{1}\left( m_{Z}\right)$ plane. Here $0\leq\protect\lambda _{2}(m_{Z})\leq0.25$ and $0\leq\protect%
\lambda _{3}(m_{Z})\leq0.25$, starting with $\vert\protect\lambda%
_{3}(m_{Z})\vert=\protect\lambda _{2}(m_{Z})$ and $\protect\lambda%
_{3}(m_{Z})=\protect\lambda _{4}(m_{Z})/2$. Red lines are the remaining
contours between $\protect\mu =10^{3}$ and $10^{19}$ GeV. Dashed line
indicates the experimental value for the ratio in $\protect\lambda %
_{1}(m_{Z})$ for a Higgs with a mass near to $125$ GeV \cite{Aad}. For red contours initial
points mark the final zone of instability scenario. Gray curve encloses the region compatible
with the strongest unitarity bound given by the eigenvalue $\boldsymbol{\Lambda} _{00}^{even+}$ in Eq. (\ref{eigen4})
.}
\label{fig:U(1)L1L4C}
\end{figure}

With the general condition $\lambda _{3}\left( \mu \right) +\lambda
_{4}\left( \mu \right) -|\lambda _{5}\left( \mu \right) |>-\sqrt{\lambda
_{1}\left( \mu \right) \lambda _{2}\left( \mu \right) }$ at specific
energies, we get the contours for $\lambda _{4}\left( m_{Z}\right) =\left(
m_{A^{0}}^{2}+m_{H^{0}}^{2}-2m_{H^{\pm }}^{2}\right) /v^{2}$ vs $\lambda
_{1}\left( m_{Z}\right) =m_{h^{0}}^{2}/v^{2}$ and $\lambda _{4}\left(
m_{Z}\right) =\left( m_{A^{0}}^{2}+m_{H^{0}}^{2}-2m_{H^{\pm }}^{2}\right)
/v^{2}$ vs $\lambda _{5}\left( m_{Z}\right) =\left(
m_{H^{0}}^{2}-m_{A^{0}}^{2}\right) /v^{2}$. For all phase diagrams for stability-instability (blue and light-blue areas respectively), we have also taken two particular backgrounds at $\mu=10^{3}$ GeV (Left-panels) and $\mu=10^{19}$ GeV (Right panels), and analyzing as critical zones (or criticality in our context) evolves with energy scales between these backgrounds.

\begin{figure}[htp]
\centering
\includegraphics[scale=0.18]{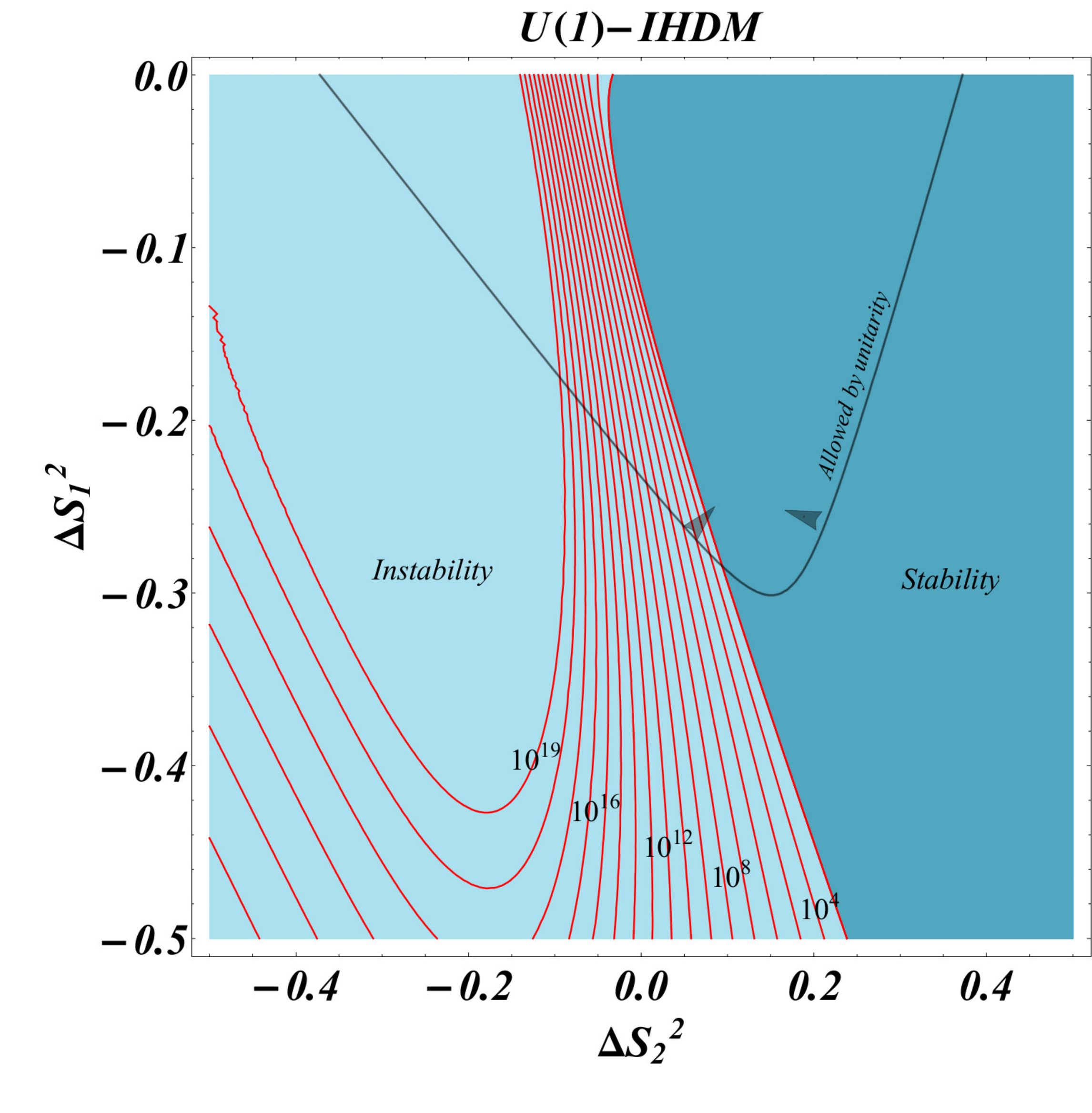} %
\includegraphics[scale=0.18]{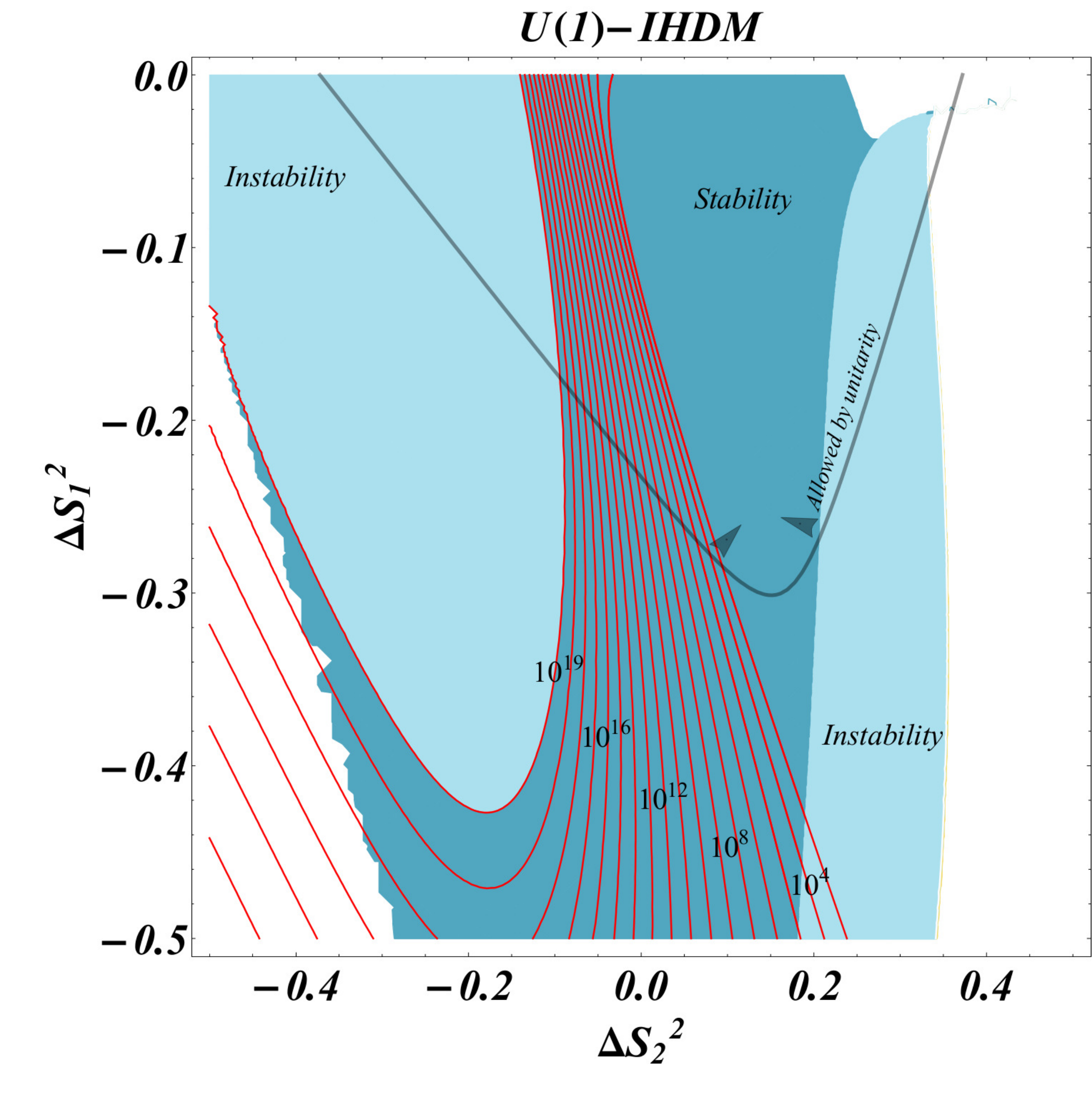} \vspace{0.6cm}
\caption{Phase diagram with evolution of stability and instability contours
from $\protect\mu =10^{3}$ GeV (\textbf{Background-Left}) up to $\protect\mu =10^{19}$ GeV
(\textbf{Background-Right}) in the $ \Delta S_{1}^{2}$ versus $\Delta S_{2}^{2}$ plane. Red lines show the evolution of the remaining
contours between $\protect\mu =10^{3}$ and $10^{19}$ GeV.  . Here 
$0\leq\protect\lambda _{2}(m_{Z})\leq0.25$, starting with $\protect\lambda _{2}(m_{Z})=\vert \protect\lambda%
_{4}(m_{Z})\vert/2$. Gray
curve encloses region compatible with the strongest unitarity bound given by eigenvalue $\boldsymbol{\Lambda} _{00}^{even+}$.}
\label{fig:U(1)L3L4}
\end{figure}

For instance in the $U(1)$ case,
contours in the $\lambda _{1}\left( m_{Z}\right) -\lambda _{4}\left(
m_{Z}\right) $ plane are depicted in Fig. \ref{fig:U(1)L1L4C}. Here
evolution of contours of stability and instability are considered in the
scales between $10^{3}$ GeV and $10^{19}$ GeV for sundry values of $%
\lambda _{3}\left( m_{Z}\right) $ and $\lambda _{2}\left( m_{Z}\right) $. Similarly, for the $U(1)$-case, we consider in (Fig. \ref{fig:U(1)L3L4}) the $%
\lambda_{3}(m_{Z})-\lambda_{4}(m_{Z})$ plane, which yields vacuum analysis
for splittings $m_{H^{\pm}}^{2}-m_{22}^{2}$ and $m_{A^{0},H^{0}}^{2}-m_{H^{%
\pm}}^{2}$ for different energy regimes. Contours have taken values around of the central ones for fermion and boson particles in the current phenomenological analyses considered in \cite{PDG,Top}. We take as our input parameters $m_{t}=173.34$ GeV, $m_{b}=4.2$ GeV, $m_{h^{0}}=125.04$ GeV, $m_{W}=80.36$ GeV, $m_{Z}=91.18$ GeV \cite{PDG}. 

\begin{figure}[tph]
\centering\includegraphics[scale=0.18]{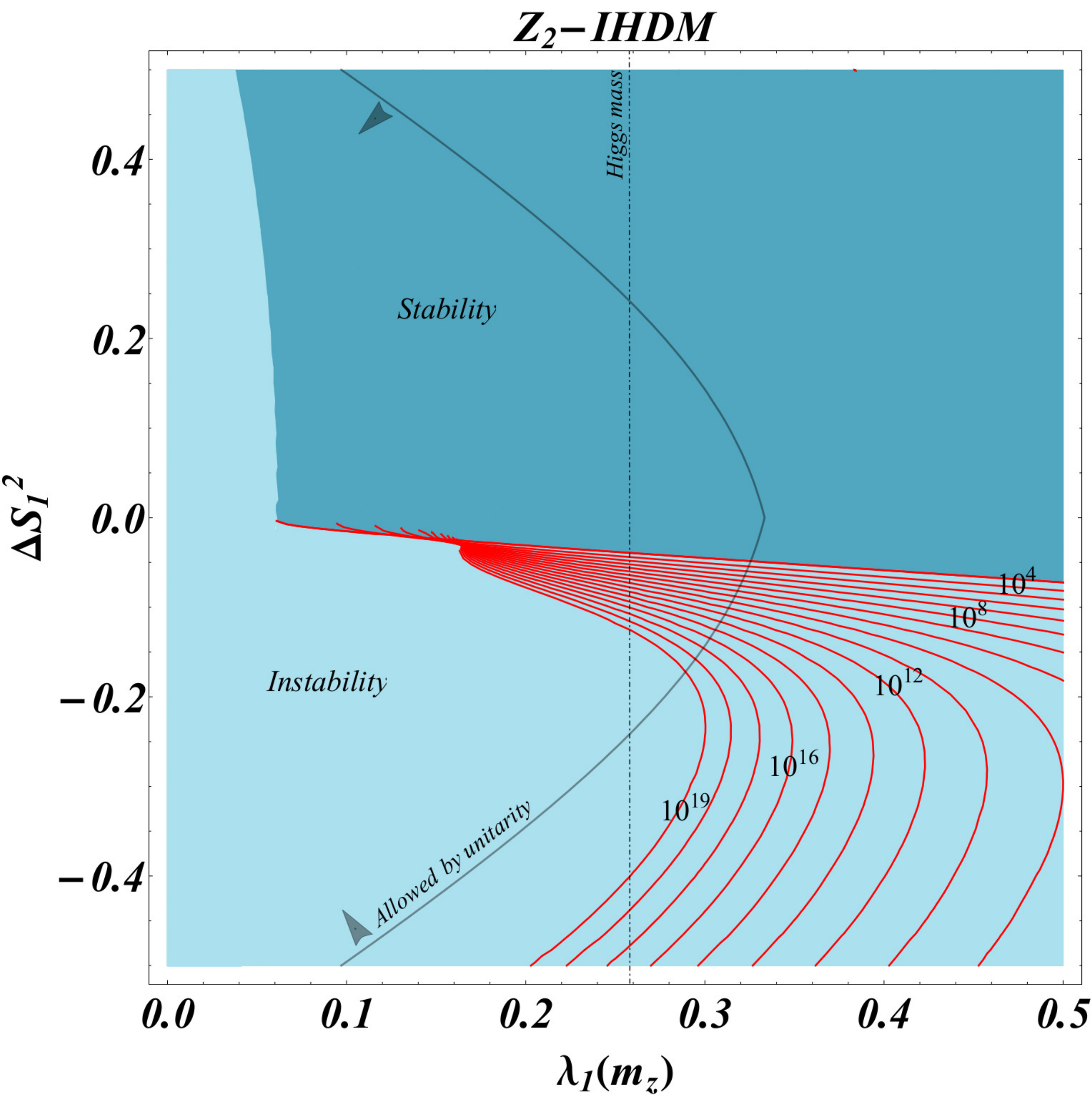} %
\includegraphics[scale=0.18]{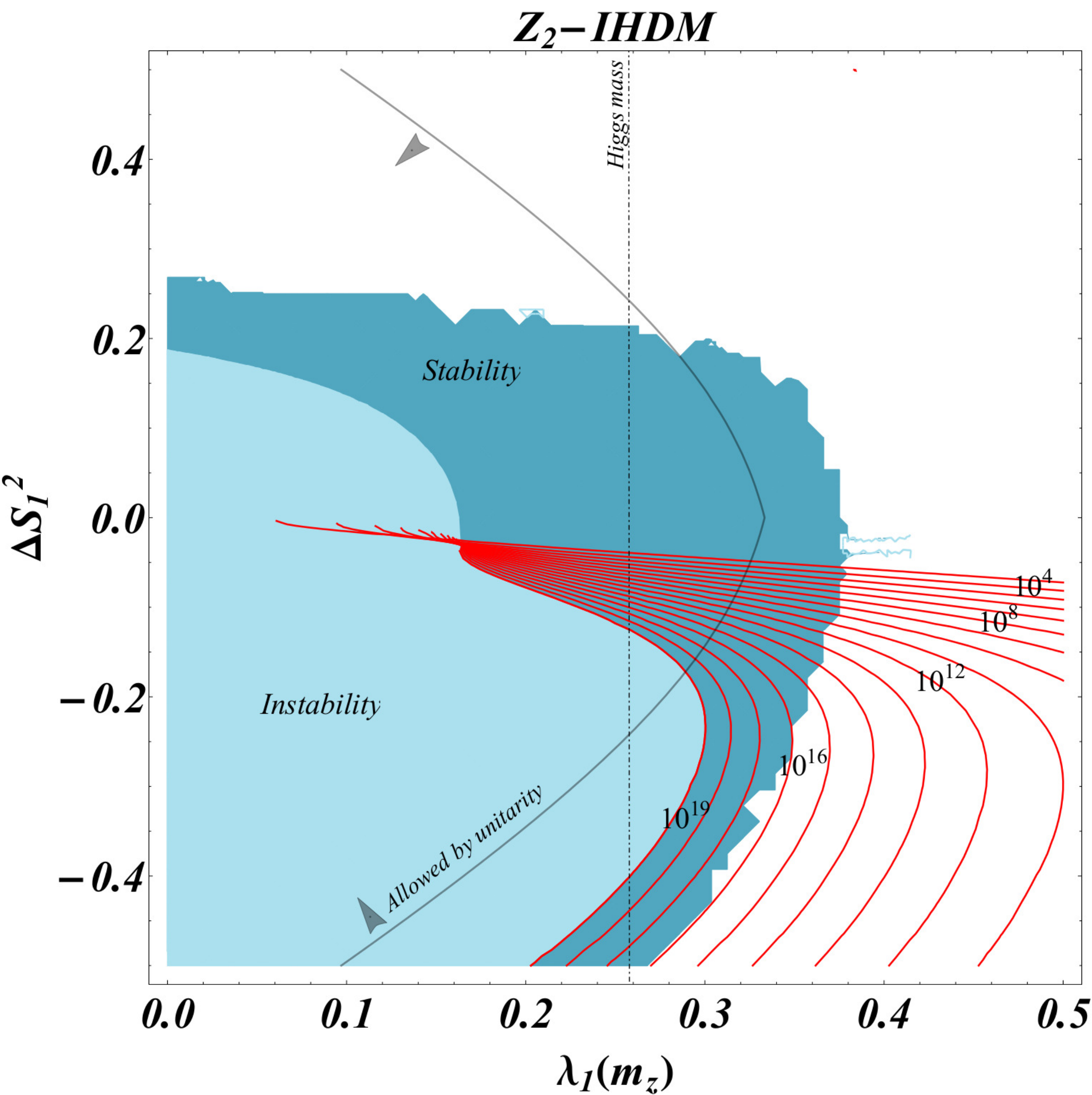} \vspace{0.6cm}
\caption{Phase diagrams with the evolution of contours from $\protect\mu %
=10^{3}$ GeV (\textbf{Background-Left}) up to $\protect\mu =10^{19}$ GeV (\textbf{Background-Right}) in the $\Delta S_{1}^{2}$ versus $\protect\lambda _{1}\left( m_{Z}\right)$ 
plane. Here $0\leq\protect\lambda _{2}(m_{Z})\leq0.25$ and $-0.25\leq\protect%
\lambda _{3,4}(m_{Z})\leq0.25$, starting with $\protect\lambda _{3,4}(m_{Z})=%
\protect\lambda _{5}(m_{Z})/2$ and $\protect\lambda _{34}(m_{Z})=\vert 
\protect\lambda_{2}(m_{Z})\vert$. Red lines are the remaining contours between $%
\protect\mu =10^{3}$ and $10^{19}$ GeV. Dashed line indicates the
experimental value for the ratio in $\protect\lambda _{1}(m_{Z})$ for a
Higgs with a mass near to $125$ GeV \cite{Aad}. Gray curve encloses region compatible with the 
strongest unitarity bound given by the eigenvalue  $\boldsymbol{\Lambda}_{00}^{even+}$.}
\label{fig:L4L1}
\end{figure}

There are regions phenomenologically relevant since they could be easily
recognizable by exclusion zones for observed or new resonances. For
instance, in the $U(1)-$model, relevant regimes are: (a) the scenario with $A^{0}-$axion appearance ($%
m_{H^{\pm }}^{2}=-\lambda _{4}v_{1}^{2}/2$) or (b) the limit for the \emph{compressed} regime with triply degenerate
scalars (i.e $m_{H^{0}}=m_{A^{0}}=m_{H^{\pm }}$). The latter occurs when $\lambda
_{4}\left( m_{Z}\right) =0.$ The first regime would also have as a consequence 
$m_{H^{0}}=0,$ which is phenomenologically unwanted and from theoretical
point of view this limit is forbidden by the model foundations. Another
important region corresponds to identify  $m_{h^{0}}$ with the
experimental resonance in the mass range of $125.04\pm 0.64$ GeV.
The theoretical framework for this assumption is the alignment regime \cite{Carena,Bernon}, which is satisfied trivially in the inert 2HDM. The zone
consistent with vacuum stability and the value $\lambda _{1}\left(
m_{Z}\right) \simeq 0.258$ (for central value of Higgs mass) in Fig. \ref%
{fig:U(1)L1L4C} and \ref{fig:L4L1} will be given as a dashed line crossing the respective
parameter space.

Constraint $\lambda _{3}\left( \mu \right) +\lambda _{4}\left( \mu \right)
-\left\vert \lambda _{5}\left( \mu \right) \right\vert >-\sqrt{\lambda
_{1}\left( \mu \right) \lambda _{2}\left( \mu \right) }$ contains the positivity conditions $\lambda
_{1}\left( \mu \right) >0$ and $\lambda _{2}\left( \mu \right) >0$
 in an independent form, because of the well
defined behavior of the root square. The RGE of $\lambda _{2}\left( \mu \right) $
is not widely relevant, because the Yukawa structure leads to an
evolution which does not involve strong sources of instabilities. Hence positivity of this product correspond to ensure positivity of $\lambda _{1}$. This correspondence between conditions can be seen in the 
contours through regions for small values of $\lambda _{1}\left(
m_{Z}\right) ,$ which will not be relevant since these regimes are located apart
from phenomenological identification of $m_{h^{0}}$.

\begin{figure}[tph]
\centering\includegraphics[scale=0.18]{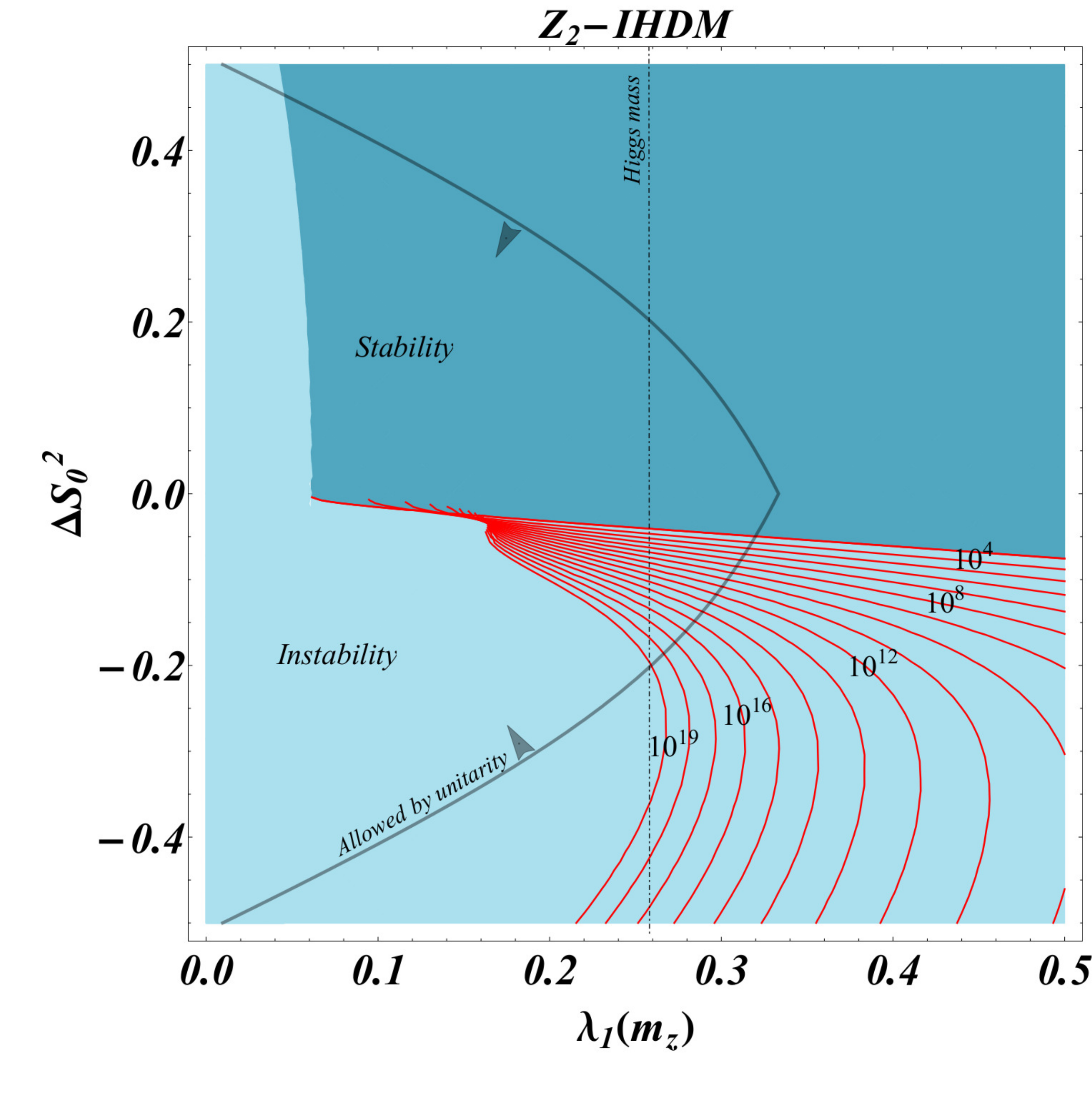} %
\includegraphics[scale=0.18]{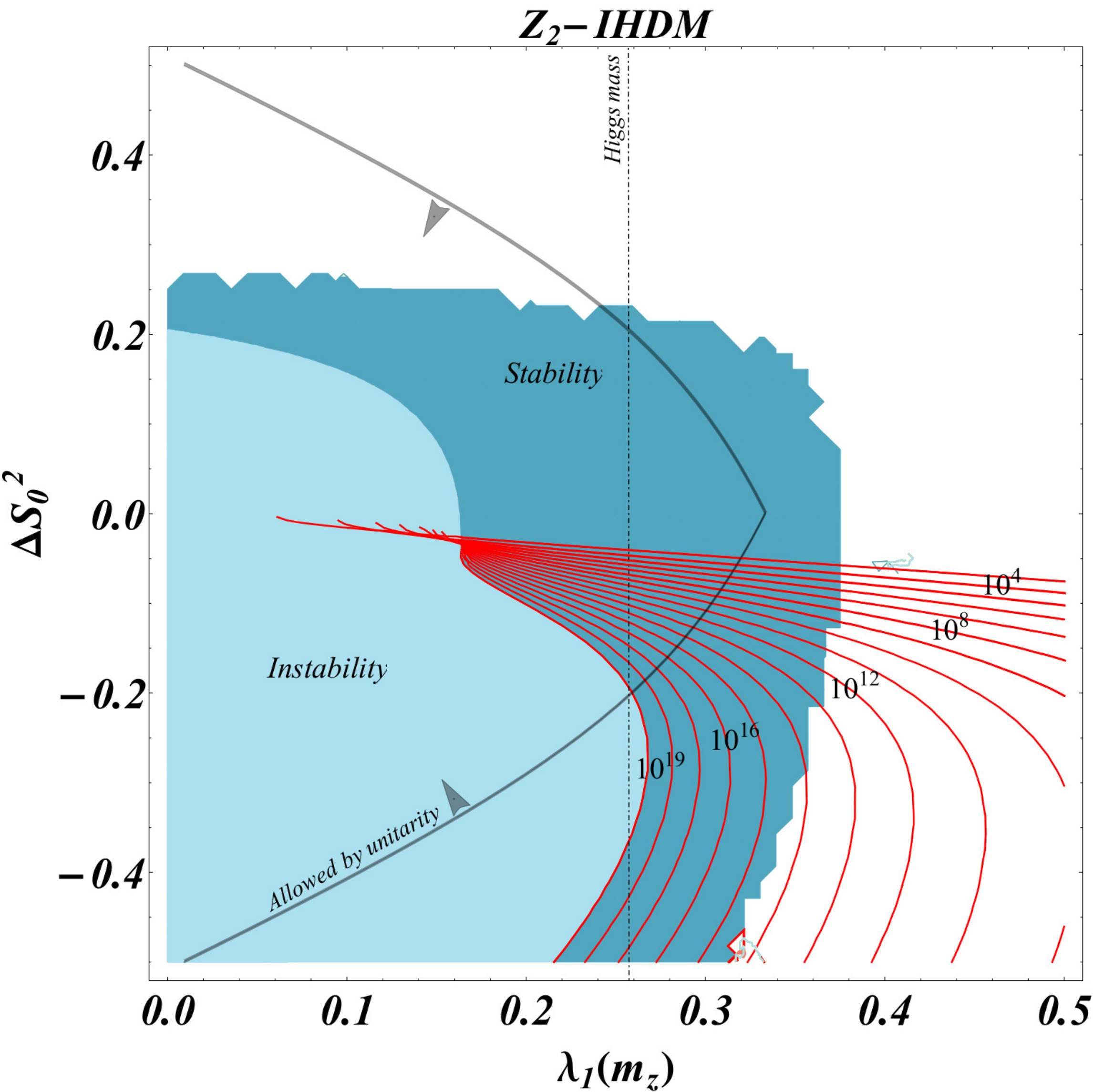} \vspace{0.6cm}
\caption{Phase diagrams with the evolution of contours from $\protect\mu %
=10^{3}$ GeV (\textbf{Background-Left}) up to $\protect\mu =10^{19}$ GeV (\textbf{Background-Right}) in the $\Delta S_{0}^{2}$ versus $\protect%
\lambda _{1}\left( m_{Z}\right)$ 
plane. Here $0\leq\protect\lambda _{2}(m_{Z})\leq0.25$ and $-0.25\leq\protect%
\lambda _{3,4}(m_{Z})\leq0.25$, starting with $\protect\lambda _{3,4}(m_{Z})=%
\protect\lambda _{5}(m_{Z})/2$ and $\protect\lambda _{34}(m_{Z})=\vert 
\protect\lambda_{2}(m_{Z})\vert$. Red lines are the remaining contours between $%
\protect\mu =10^{3}$ and $10^{19}$ GeV. Dashed line indicates the
experimental value for the ratio in $\protect\lambda _{1}(m_{Z})$ for a
Higgs with a mass near to $125$ GeV \cite{Aad}. Gray curve encloses the region compatible with the 
strongest unitarity bound given by the eigenvalue $\boldsymbol{\Lambda}_{00}^{even+}$.}
\label{fig:L5L1}
\end{figure}

\begin{figure}[tph]
\centering\includegraphics[scale=0.18]{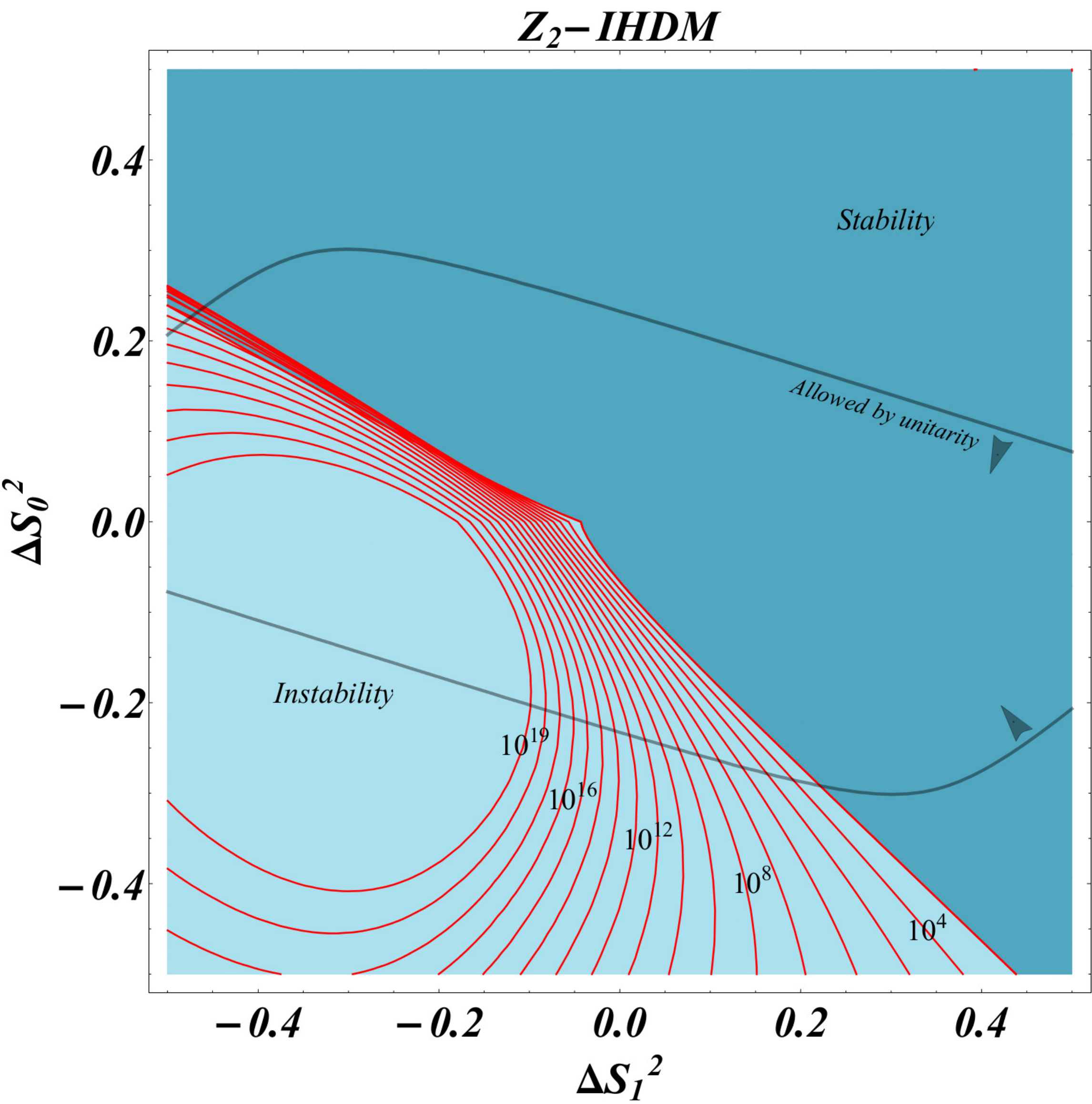} %
\includegraphics[scale=0.18]{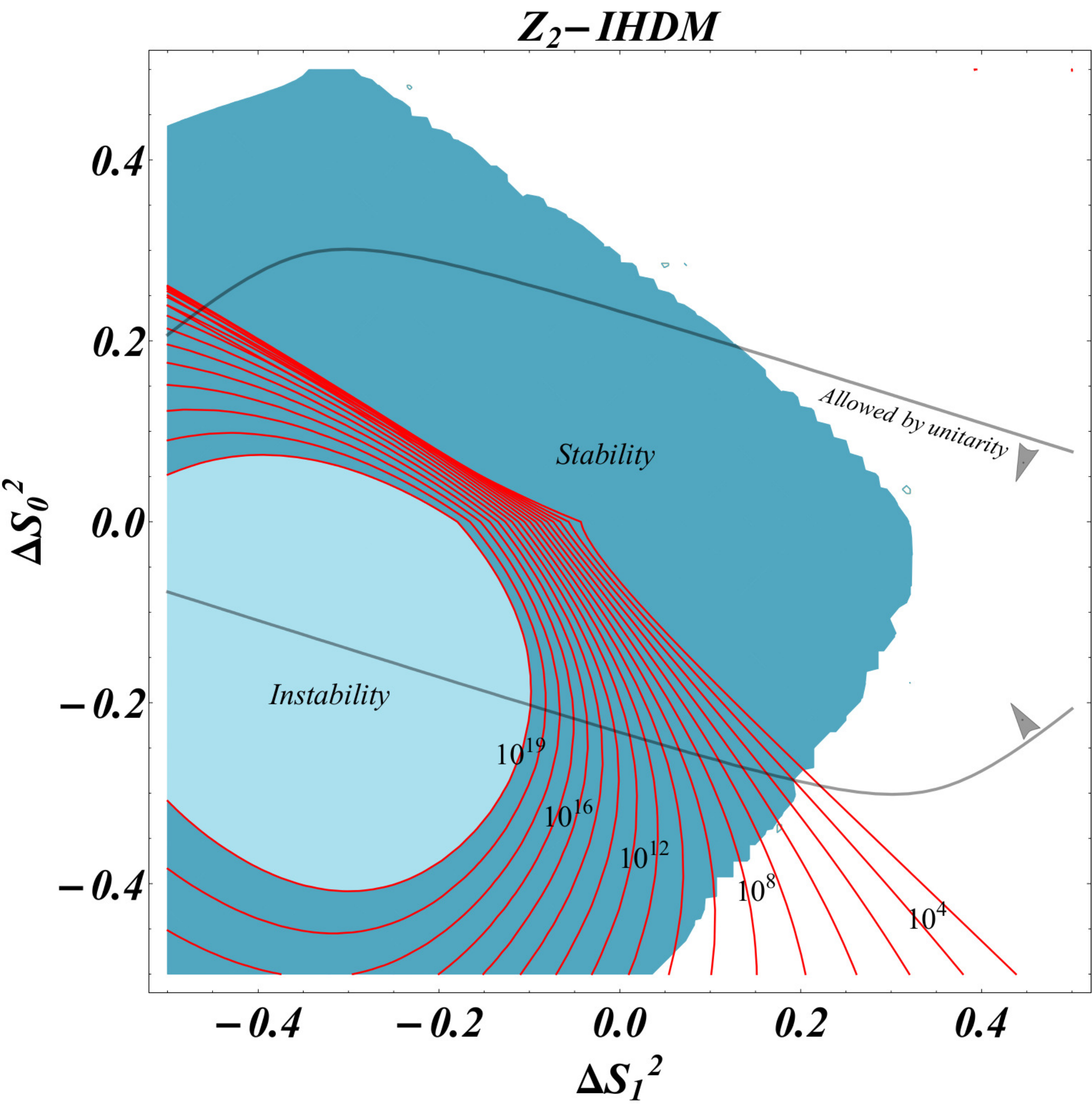} \vspace{0.6cm}
\caption{Phase diagram with evolution of stability and instability contours
from $\protect\mu =10^{3}$ GeV (\textbf{Background-Left}) up to $\protect\mu =10^{19}$ GeV
(\textbf{Background-Right}) in the $\Delta S_{0}^{2}$ versus $\Delta S_{1}^{2}$ plane. Here 
$0\leq\protect\lambda _{2}(m_{Z})\leq0.25$ and $-0.25\leq\protect\lambda %
_{3}(m_{Z})\leq0.25$, starting with $\protect\lambda _{3}(m_{Z})=\protect%
\lambda _{5}(m_{Z})/2$ and $\protect\lambda _{2}(m_{Z})=\vert \protect\lambda%
_{3}(m_{Z})\vert$. Red lines show the evolution of the remaining contours between $%
\protect\mu =10^{3}$ and $10^{19}$ GeV.  Gray curves enclose the region
compatible with the strongest unitarity bound of the eigenvalue $\boldsymbol{\Lambda}
_{00}^{even}$.}
\label{fig:L5L4}
\end{figure}

Correspondingly, contours in the $Z_{2}$-model are shown in Figs. \ref{fig:L4L1}-\ref{fig:L5L4}, from which we can study stability behavior in $%
\lambda_{1}(m_{Z})-\lambda_{4}(m_{Z})$ and $\lambda_{1}(m_{Z})-%
\lambda_{4}(m_{Z})$ planes. As in the $U(1)$ case, dashed line indicates $%
m_{h^{0}}$ identification with Higgs-like scalar observed in LHC. In Fig. %
\ref{fig:L5L4} we show the variation of vacuum stability and instability zones
for $\lambda _{3}\left( \mu \right) +\lambda _{4}\left( \mu \right)
-\left\vert \lambda _{5}\left( \mu \right) \right\vert >-\sqrt{\lambda
_{1}\left( \mu \right) \lambda _{2}\left( \mu \right) }$ contour with
respect to the energy scale in $\lambda _{5}\left( m_{Z}\right) -\lambda _{4}\left(
m_{Z}\right) $ plane, which leads to determine the mass splittings
influence in vacuum stability for the $Z_{2}$ case.

Our phase diagrams lead us to verify some limits for the perturbative validity of the both models ($Z_{2}$ -$U(1)$) in the field space. It can be seen due to solutions for RGEs present possible Landau poles. These non-perturbative zones are identified with white areas, as it is shown on the right side of Figs. \ref{fig:U(1)L1L4C}-\ref{fig:L5L4} for the background of $\mu=10^{19}$ GeV.

\subsection{Implications of vacuum behavior in $Z_{2}-$2HDM and $U\left(
1\right)-$2HDM}
\label{sec:AR}

In the SM, the positivity of the scalar boson mass-squared and
bounded from below potential implies that $\lambda >0.$ To ensure vacuum
stability for all scales up to $\mu_{I} $, one must have $\lambda \left( \mu
\right) >0$ for all $\mu $ between $m_{Z}$ and $\mu_{I} .$ Similarly, to
ensure vacuum stability in the 2HDM up to $\mu_{I}$, effective Higgs potential must require that
all of the five constraints be valid up to $\mu_{I} $. At one loop level,
this can be rendered as a threshold effect for SM vacuum. If the condition $%
\lambda _{1}\left( \mu \right) >0$ or $\lambda _{2}\left( \mu \right) >0$ is
violated, the potential will be unstable in the $\Phi _{1}$ or $\Phi _{2}$ direction respectively. These threshold corrections at one loop increase the Higgs potential stability by the introduction of new fields and couplings among them, which
in the SM is lost even from scales around $\mu =10^{11}$ GeV \cite{Degrassi}. Although in this case, new physics improves vacuum stability in $\Phi_{1}$ in
particular limits compatible with the SM behavior, other directions can be affected by
the fields and couplings added to the spectrum.

In other directions of the extended field space, the statement of instability works as follows: if the conditions $\lambda _{4}\left( \mu \right) +\lambda _{5}\left(
\mu \right) <0,$ $\lambda _{3}\left( \mu \right) +\lambda _{4}\left( \mu
\right) -\left\vert \lambda _{5}\left( \mu \right) \right\vert +\sqrt{%
\lambda _{1}\left( \mu \right) \lambda _{2}\left( \mu \right) }>0$ or $\lambda _{3}\left( \mu \right) +\sqrt{\lambda _{1}\left( \mu \right)
\lambda _{2}\left( \mu \right) }>0$ are not accomplished one by one or
simultaneously, the potential will be unstable in the $\Phi _{1}$-$\Phi _{2}$
plane. At the same time, it is viable to require that all $\lambda $'s
be finite (or perturbative) up to $\Lambda $ in order to avoid possible Landau poles.

Numerical analyses start with quartic couplings defined at the
electroweak scale $\mu _{ew}=m_{Z}$. With these initial conditions, the RGEs are
integrated out to search whether one of the bounds for positivity is
violated or whether any of the couplings become non-perturbative before
reaching a $\mu_{\text{crit}}\equiv\Lambda $ (procedure established in \cite{Sher-Nie}). By sweeping different zones in the parameter space, it is possible to describe contours as a function of scalar mass splittings.  The contours built up, interpreted as phase diagrams, yield information about how instabilities arise in the Higgs potential at an energy scale and a field-space direction given.

Minimality principle and vacuum relations could be studied by some
limits between the SM and the inert-2HDM. For instance, the parameter space compatible with $h^{0}$ emulating to SM-Higgs boson is non-suppressed even at Planck scales for positive values of $\lambda _{5},\lambda _{4},\lambda _{3}$ as it is shown in vacuum stability
analyses. This regime implies that, for $\lambda _{1}\left( m_{Z}\right) $
identified with the Higgs mass and with initial conditions, $%
\lambda _{5}\left( m_{Z}\right) $ must be positive and whose inferior limit close to $0.0.$. The limit superior belongs in $\lambda_{5}(m_{Z})\simeq0.2$, which is also compatible with unitarity perturbative bounds.  In
the $Z_{2}$ case, this stable zone is consistent with a spectrum where $%
m_{H^{0}}>m_{A^{0}}$ for all energy scales. Nonetheless, this region will be suppressed by condition $\lambda_{4}>\lambda_{5}$.

Vacuum stability and perturbative unitarity bounds are compatible with $\Delta
S_{0}^{2}>0$ (in a wide zone), while there also exist a reduced zone where $\Delta
S_{0}^{2}<0$ is allowed by both analyses. The last result is compatible with the tree level analysis where $\lambda_{4}>\lambda_{5}$. From plane $\lambda _{4}\left( m_{Z}\right)
-\lambda _{1}\left( m_{Z}\right) ,$ a similar restriction over $\Delta
S_{1}^{2}$ implies $2m_{H^{\pm }}^{2}<m_{H^{0}}^{2}+m_{A^{0}}^{2}.$
Compatibility among $\lambda_{4}+\lambda_{5}<0$ $\lambda_{4}-\lambda_{5}>0$
and vacuum stability scenario gives an advantage for regions where $%
\lambda_{4}>0$ and $\lambda_{5}<0$, with small splittings. The last fact is a radical difference between both models, because in the $U(1)$-2HDM and to avoid vacuum configurations with charge violation, the model demands $%
\lambda_{4}<0$.

Non-perturbative values are driven out for $\lambda _{1,2}\left(
m_{Z}\right) \sim 0.25$ (even incompatible with perturbative
unitarity) and $\lambda _{3}\left( m_{Z}\right) \sim -0.35$ and $\lambda
_{3,4,5}\left( m_{Z}\right) \sim 0.25,$ being determined by regions where
 numerical solutions of RGEs were finite. As it was pointed out, these
non-perturbative zones are also strongly disfavored by unitarity constraints of scalar scattering processes.

\section{Tree level contours for metastability analyses}
\label{sec:metastabilityanalyses}

\begin{figure}
\centering
\includegraphics[scale=0.18]{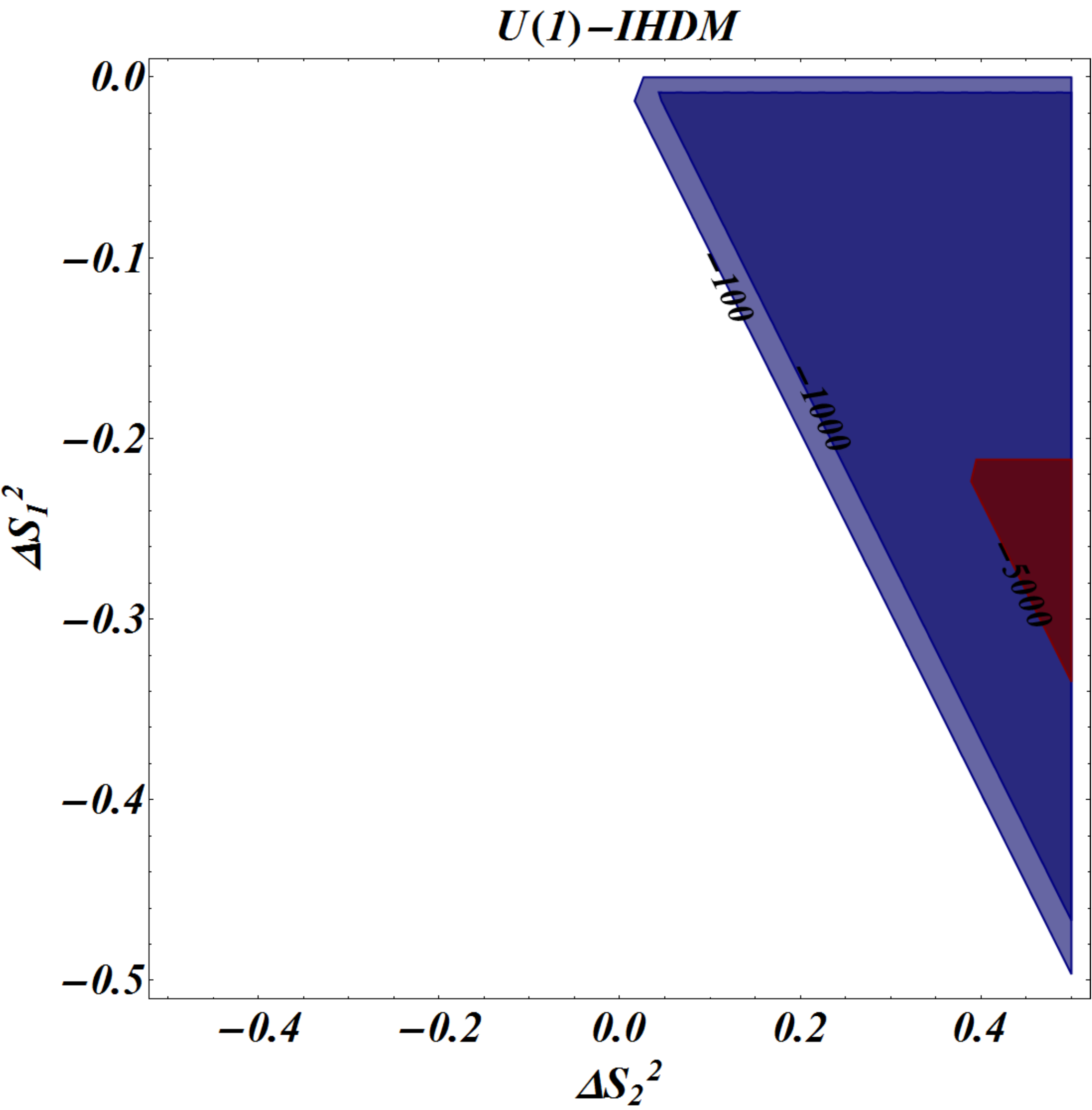}
\includegraphics[scale=0.18]{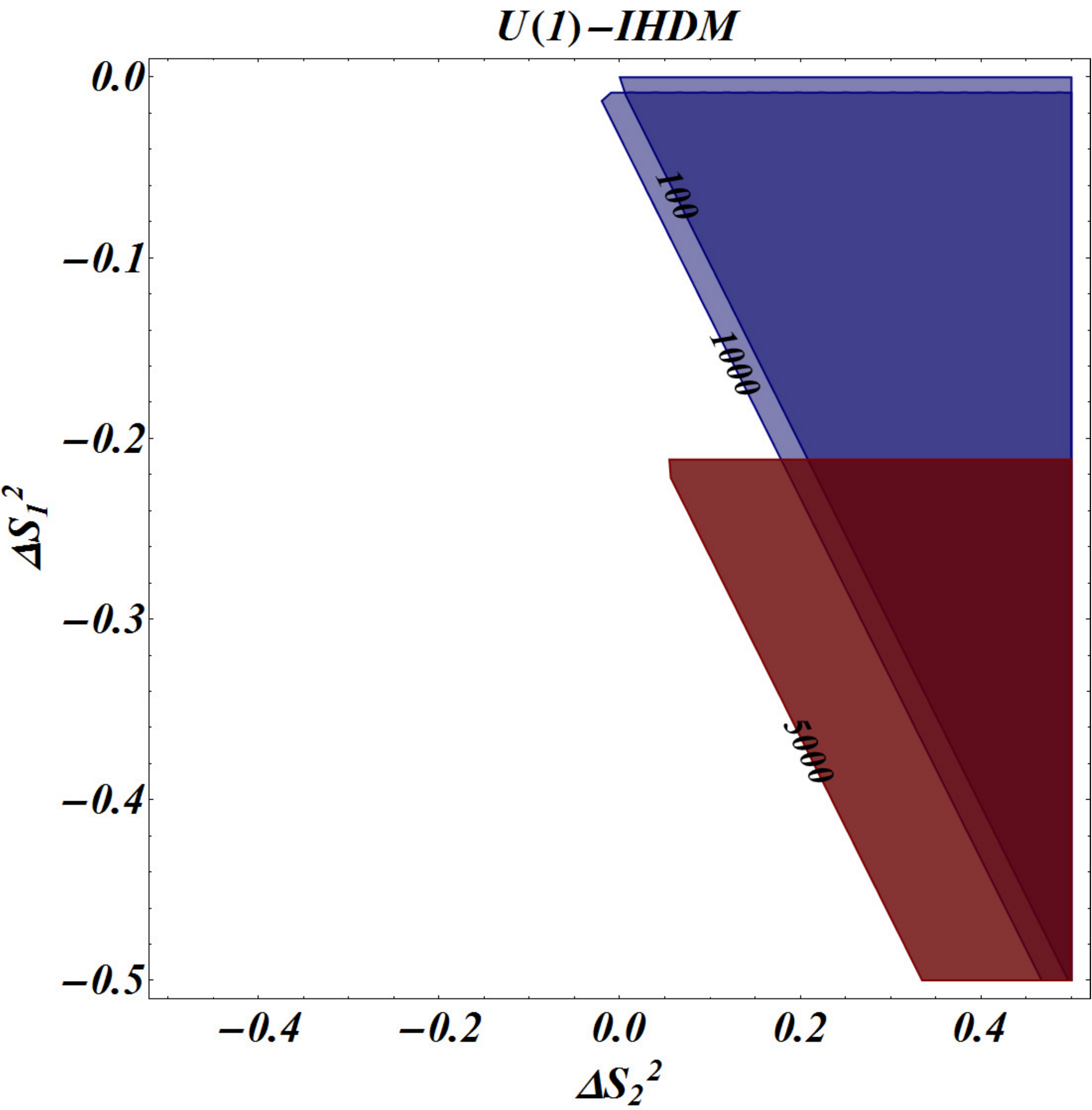}
\vspace{0.5cm}
 \caption{Electroweak global minimum for $\Delta S_{1}^{2}-\Delta S_{0}^{2}$ zones in different values of (\textbf{Left}) $m_{22}^{2}\left(\text{GeV}^{2}\right)=-100,-1000,-5000$  and (\textbf{Right}) $m_{22}^{2}\left(\text{GeV}^{2}\right)=100,1000,5000$. Here 
$0\leq\protect\lambda _{2}\leq0.25$, starting with $\protect\lambda _{2}=\vert \protect\lambda%
_{4}\vert/2$.  }
 \label{fig:metastabilityU1}
\end{figure}

To search the compatibility between splittings allowed by vacuum analyses and the presence of a global minimum in these scenarios, we take into account the restrictions obtained in Eqs (\ref{Meta1})-(\ref{Meta3}). 
For instance, in the $U(1)$-model, we evaluate the metastability constraints over $\Delta S_{1}^{2}-\Delta S_{2}^{2}$ plane in Fig. \ref{fig:metastabilityU1}. We see as negative and positive values of $m_{22}^{2}$ favor positive zones for $\Delta S_{2}^{2}$, relating to $m_{H^{\pm}}^{2}>m_{22}^{2}$ hierarchy. These compatible zones are larger for $\vert m_{22}^{2}\vert<5000$ GeV$^{2}$. Particularly values $\vert m_{22}^{2}\vert>7500$ GeV$^{2}$ will enter in higher values of $\Delta S_{2}^{2}$, which are related to non-perturbative or unstable zones in the Higgs potential. Compressed models with an approximate degeneracy between $m_{H^{\pm}}=m_{A^{0},H^{0}}$ become incompatible with a global minimum  in $\vert m_{22}^{2}\vert>1000$ GeV$^{2}$

\begin{figure}
\centering
\includegraphics[scale=0.18]{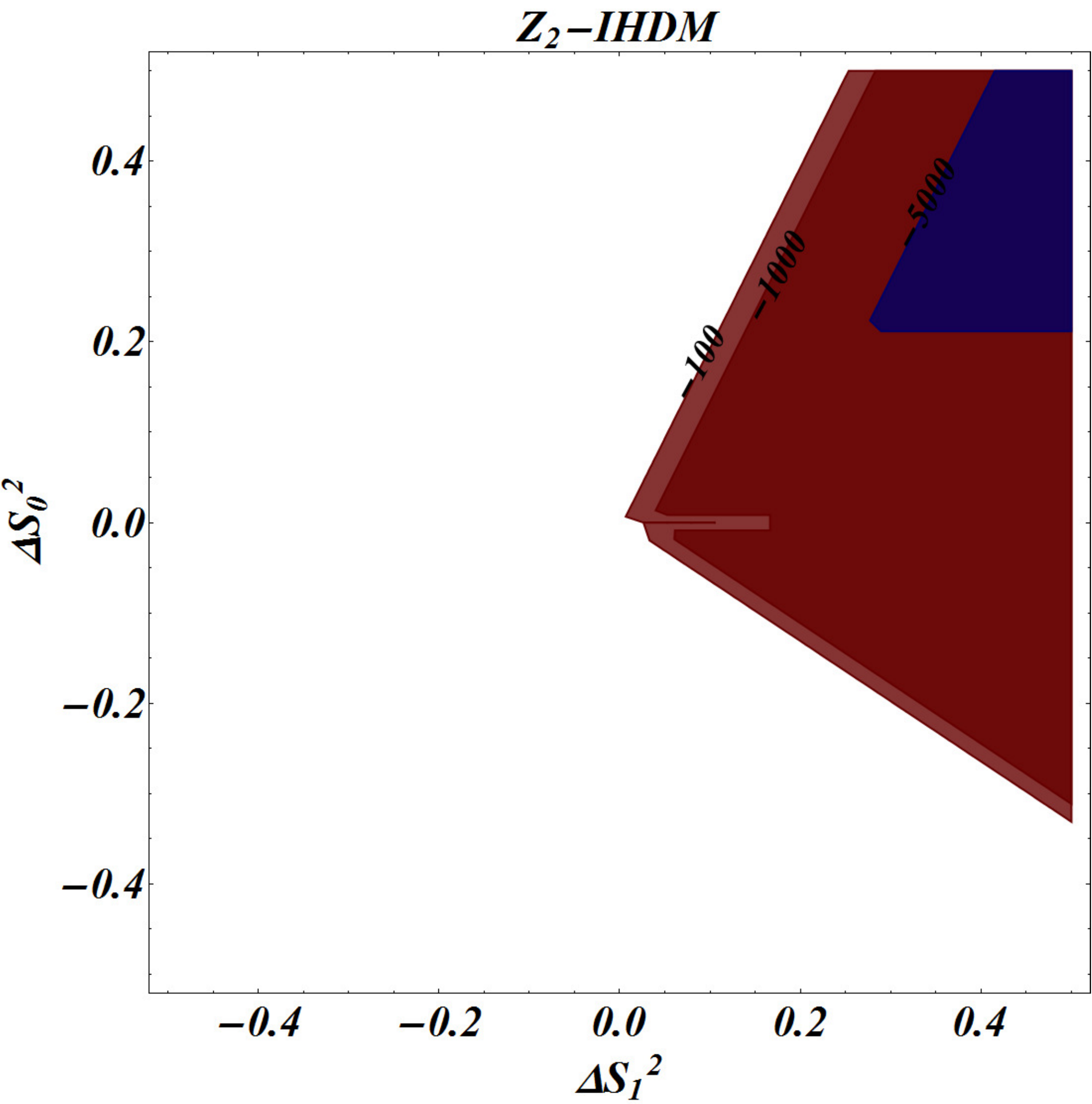}
\includegraphics[scale=0.18]{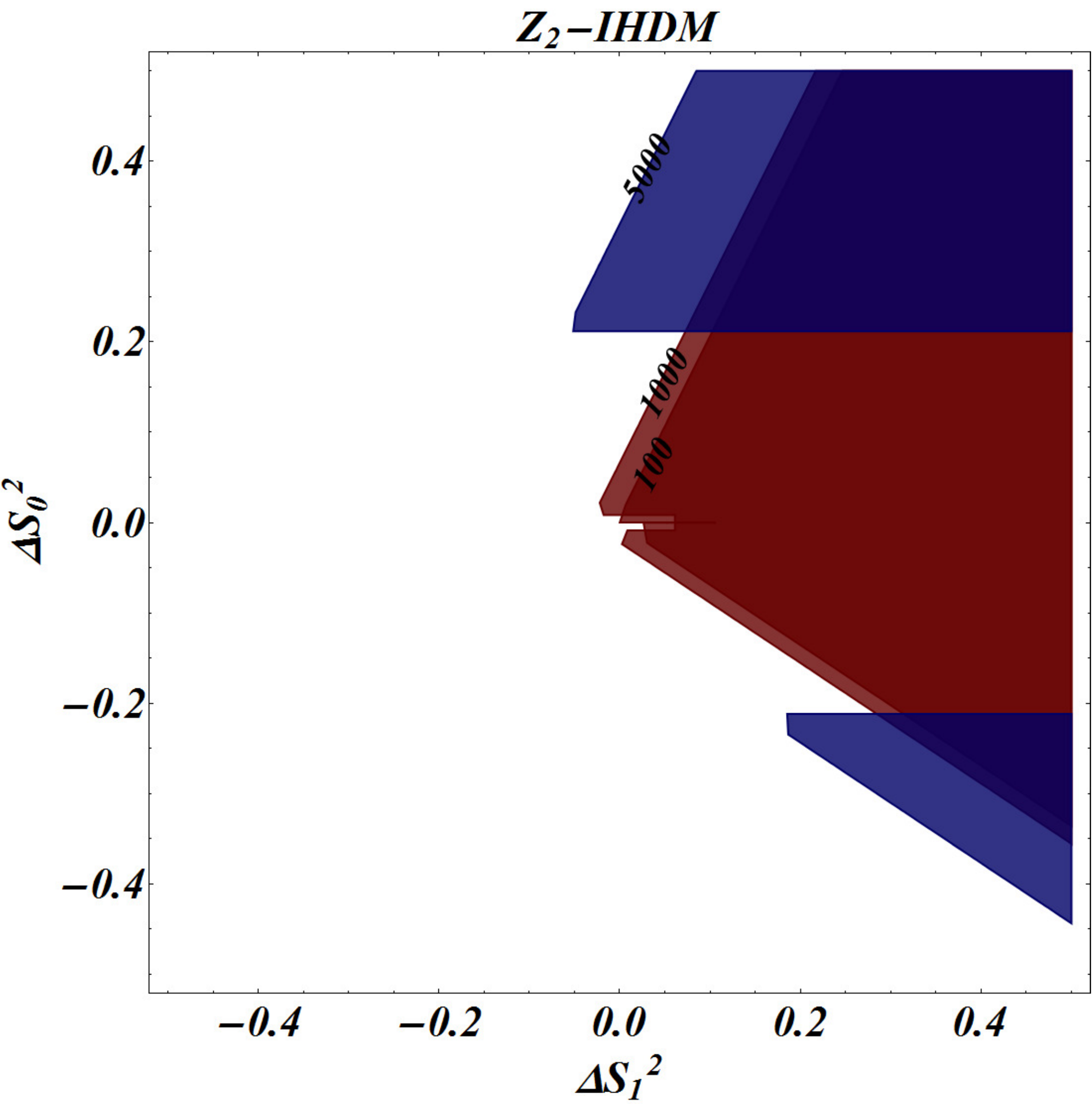}
\vspace{0.5cm}
 \caption{Electroweak global minimum for $\Delta S_{0}^{2}-\Delta S_{1}^{2}$ zones in different values of (\textbf{Left}) $m_{22}^{2}\left(\text{GeV}^{2}\right)=-100,-1000,-5000$  and (\textbf{Right}) $m_{22}^{2}\left(\text{GeV}^{2}\right)=100,1000,5000$. Here 
$0\leq\protect\lambda _{2}\leq0.25$ and $-0.25\leq\protect\lambda %
_{3}\leq0.25$, starting with $\protect\lambda _{3}=\protect%
\lambda _{5}/2$ and $\protect\lambda _{2}=\vert \protect\lambda%
_{3}\vert$. }
 \label{fig:metastabilityZ2}
\end{figure}

In the $Z_{2}$ case, in Fig. \ref{fig:metastabilityZ2} global minima zones are drawn over $\Delta S_{1}^{2}-\Delta S_{0}^{2}$ plane for different values of $m_{22}^{2}$, which generate a global minimum in positive values of $\Delta S_{1}^{2}$.  Those regimens are compatible with assumptions from conditions to avoid a charge violation minimum.
Other important point is that compressed models become incompatible with a EW-global minimum for $\vert m_{22}^{2}\vert >2500$ GeV$^{2}$. Explicitly, in $m_{22}^{2}=5000$ GeV$^{2}$ just values of $\vert \Delta S_{0}^{2}\vert >0.2$ are compatible with a global minimum structure, but incompatible with a mass degeneracy between $A^{0}$ and $H^{0}$.

\section{Oblique parameters and observables influence}

\label{sec:ST} From vacuum and metastability analyses, it is possible to make a major comparison with electroweak precision parameters since they are highly
sensitive to mass splittings \cite{429Report}. It is well known that oblique
parameters are designed to constrain models of new physics from the
electroweak precision observables. It is assumed that the effects of new
physics only appear through vacuum polarization and therefore enables us
to modify oblique parameters. Most of the effects on electroweak precision
observables can be parameterized by three gauge self-energy parameters $(S, T, U)$ introduced by Peskin and Takeuchi \cite{STU}-\cite{STU3}. Hence, the
correlation among the parameters above could be given regarding electroweak
observables and leads to analyses some precision physics, useful
to constraint phenomenology from new physics mechanisms. For instance, $S$ or $S+U$ describe new
physics contributions to neutral (charged) current processes at several
energy scales; while $T$ measures the difference between the new physics
contributions of neutral and charged current processes at low energies
(i.e., sensitive to isospin violation) close to EW cut \cite{Goebel}. Indeed, this parameter is related to the
commonly used parameter $\rho _{0}=\rho /\rho _{SM}$ through $\rho
_{0}=1/\left( 1-\alpha T\right) ;$ encoding the departure from the SM
value of $\rho _{0}=1$. By contrast, $U$ is only constrained by the $W$ boson mass and its total width. Likewise, $U$ is seldom small in new physics models, and therefore, the $STU$ parameter space can often be projected down to a two-dimensional parameter space in which the experimental constraints are easy to visualize\cite{Goebel}
\footnote{In fact, $U$ quantity is related to a dimension-eight operator, while $S$ and $T$ can be given concerning six dimension operators.}.

Constraints on the STU parameters are derived from a
fit to the precision electroweak data (more details can be found in the most
current articles \cite{EWprecision}-\cite{EWprecision5}). Besides, in the STU parameters the floating
fit values are $m_{Z}=91.1873\pm 0.0021$ GeV, $\Delta \alpha
_{had}(m_{Z}^{2})=0.02757\pm 0.00010$, and $\alpha _{s}(m_{Z}^{2})=0.1192\pm
0.0033$. The following fit results are determined from a fit for a reference
Standard Model with $m_{t,ref}=173$ GeV and $m_{H,ref}=125$ GeV and fixing $U=0$:
giving $S_{U=0}=0.06\pm 0.09$ and $T_{U=0}=0.10\pm 0.07$,
with a correlation coefficient of $+0.91$. The general procedure to measure oblique parameters relies on a global fit to the high-precision electroweak observables coming from particle collider experiments (mostly the \emph{Z}
pole data from the CERN-LEP collider) and atomic parity violation \cite%
{STU}-\cite{Gfitter}. Every step presented here would be a valuable tool to
measure the compatibility level of the vacuum behavior predictions with the
EW observables and precision tests.

Despite at this level, these computations do not distinguish among fermionic couplings, the plane of
correlations gives information about scalar states splitting and how it
could be restricted from EW measurements. Definitions of $S$ and $T$
parameters for 2HDM-Inert case read \cite{Barbieri,STIHDM}:

\begin{align}
 S_{In}&=\frac{1}{2\pi}\left[\frac{1}{6}\ln\left(\frac{m_{H^{0}}^{2}}{m_{H^{\pm}}^{2}}\right)+\frac{1}{3}\frac{m_{H^{0}}^{2}m_{A^{0}}^{2}}{\left(m_{A^{0}}^{2}-m_{H^{0}}^{2}\right)^{2}}+\frac{1}{6}\frac{m_{A^{0}}^{4}\left(m_{A^{0}}^{2}-3m_{H^{0}}^{2}\right)}{\left(m_{A^{0}}^{2}-m_{H^{0}}^{2}\right)^{3}}\ln\left(\frac{m_{A^{0}}^{2}}{m_{H^{0}}^{3}}\right)-\frac{5}{36}\right],\\
 T_{In}&=\frac{1}{32\pi\alpha^{2}v^{2}}\left[F\left(m_{H^{\pm}},m_{H^{0}}\right)+F\left(m_{H^{\pm}},m_{H^{0}}\right)-F\left(m_{A^{0}},m_{H^{0}}\right)\right],
\end{align}

with $F$ a masses symmetric function defined by

\begin{align}
 F(m_{1},m_{2})\equiv\frac{m_{1}^{2}+m_{2}^{2}}{2}-\frac{m_{1}^{2}m_{2}^{2}}{m_{1}^{2}-m_{2}^{2}}\ln\left(\frac{m_{1}^{2}}{m_{2}^{2}}\right).
\end{align}

\begin{figure}[tbp]
\centering\includegraphics[scale=0.18]{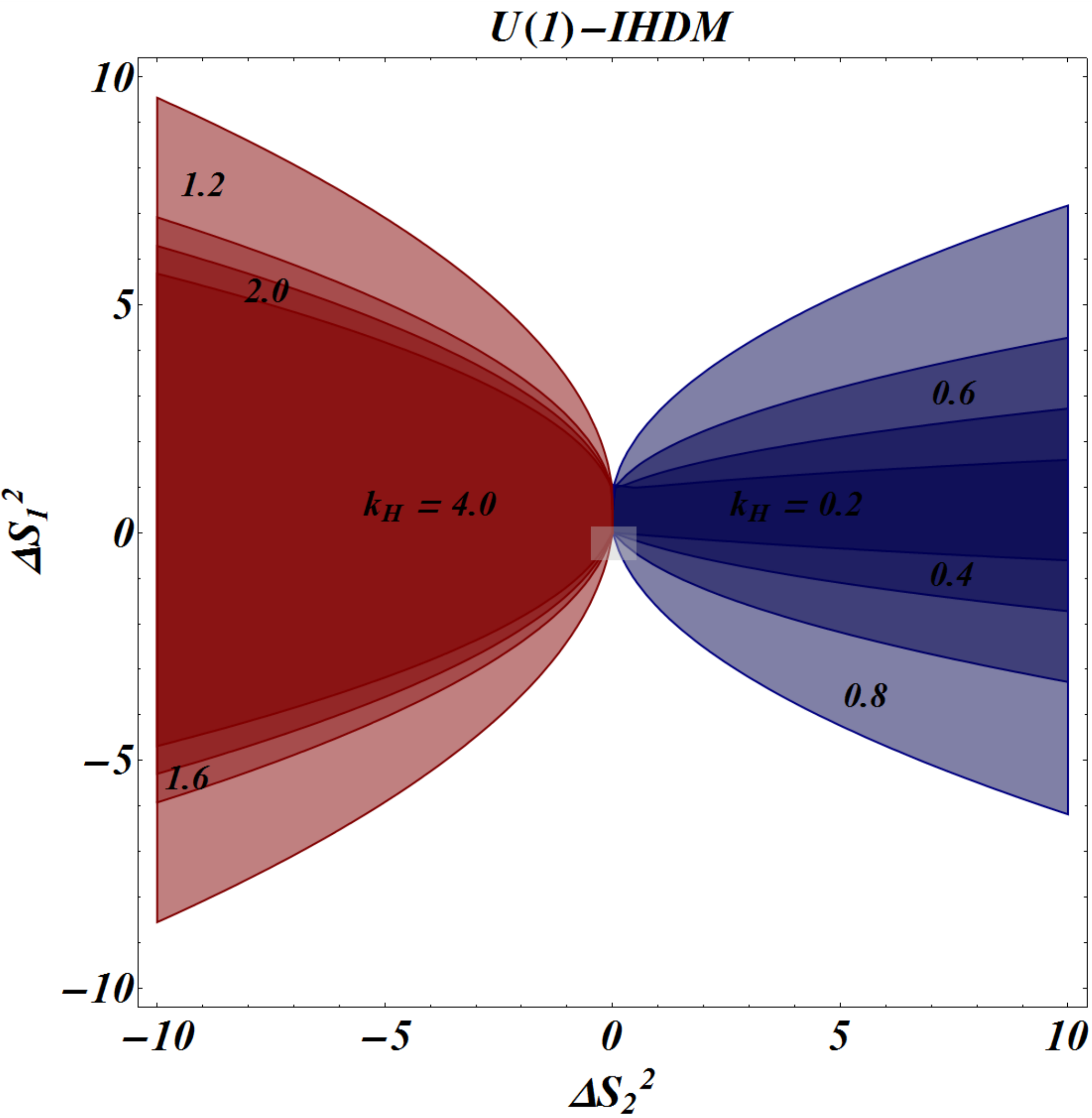} 
\includegraphics[scale=0.18]{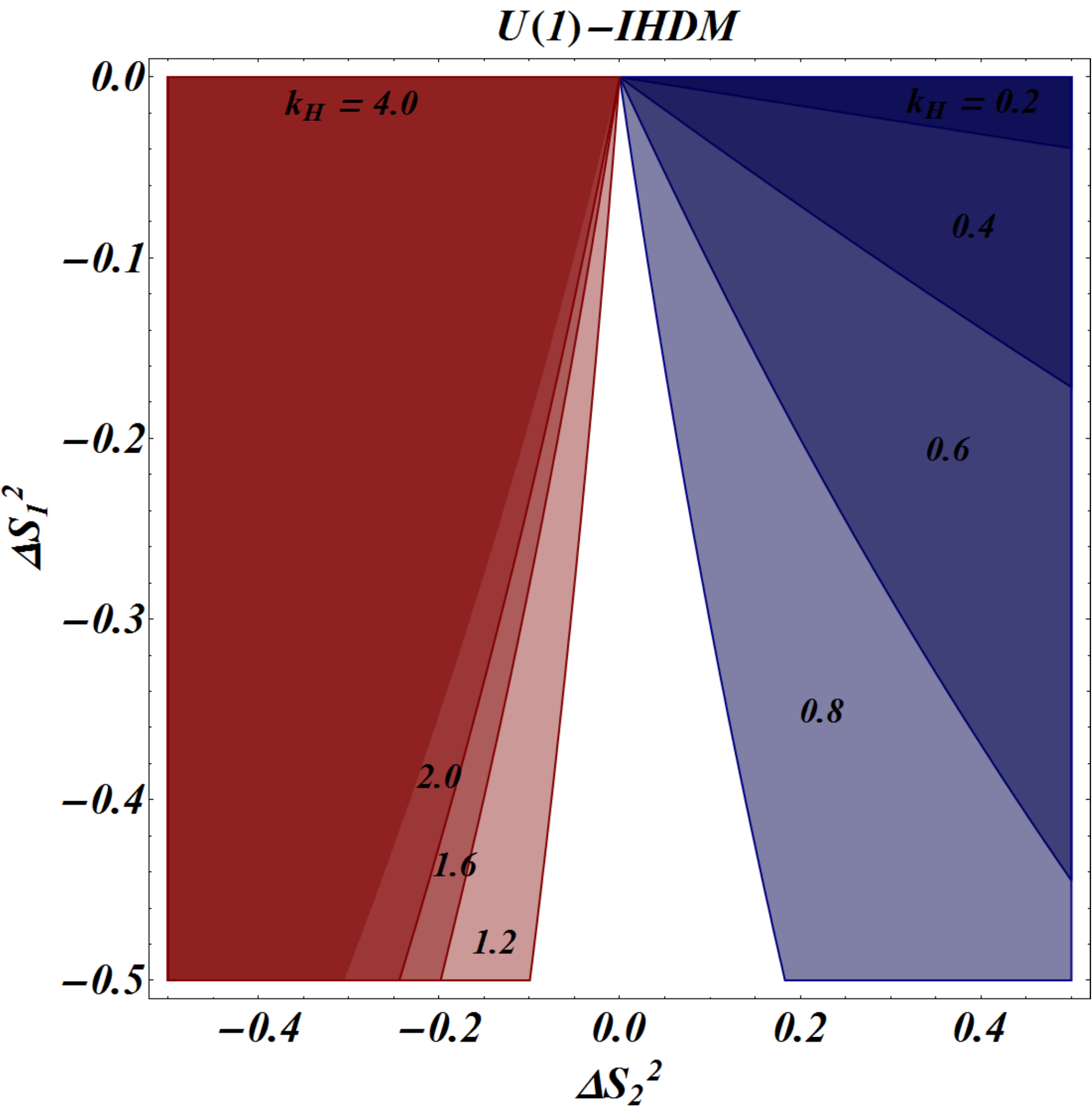}\\
\includegraphics[scale=0.18]{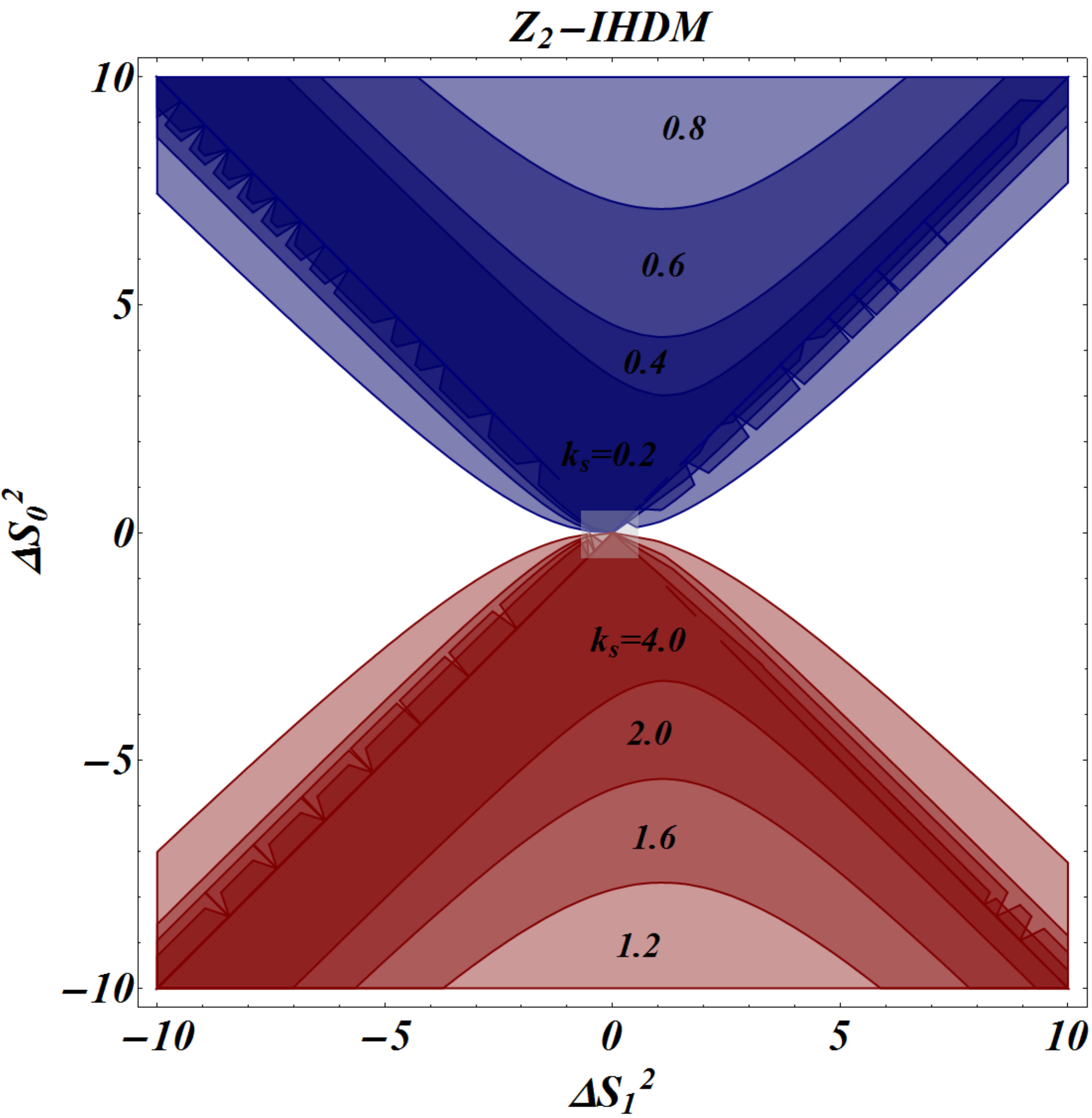} 
\includegraphics[scale=0.18]{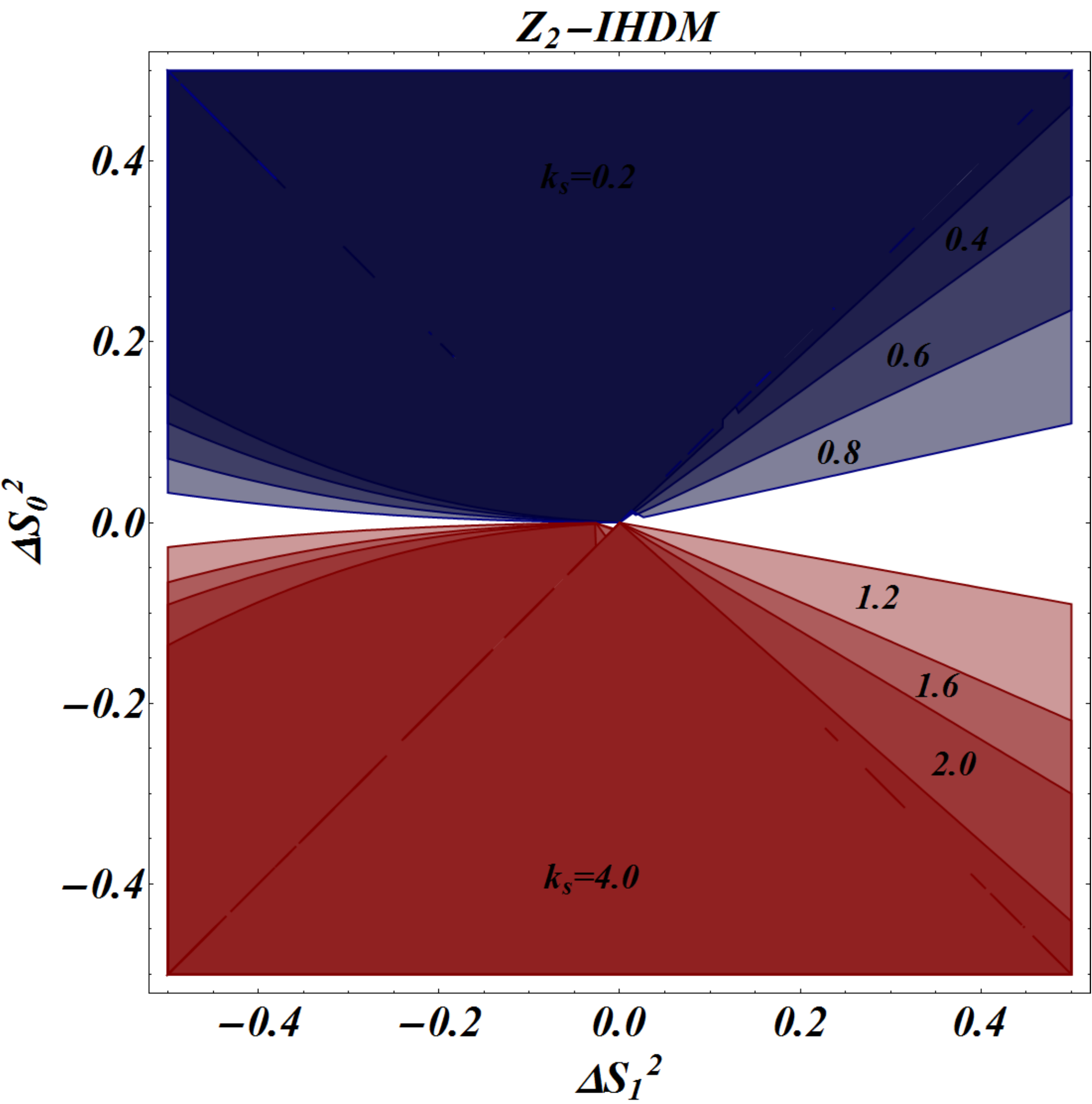}\vspace{0.6cm}
\caption{Oblique parameters in the inert-Higgs doublet model with the \emph{S%
}, \emph{T} fit results (with \emph{U} = 0) at 99 $\%$  CL  for $U(1)$ (\textbf{Up}) and $Z_{2}$ (\textbf{Down})
symmetries. The model area is obtained with the use of the mass parameter splittings
and defining $k_{H}^{2}\equiv m_{22}^{2}/m_{H^{\pm}}^{2}$ ($U(1)$ case) and $k_{s}\equiv m_{A^{0}}/m_{H^{0}}$ ($Z_{2}$ case) ratios. Plots in right side are zoomed regions compatible with vacuum stability analyses. Computations over $ST$ plane have used Mathematica module described in \cite{Deva}.}
\label{fig:OP}
\end{figure}

From the equations for \emph{S} and \emph{T} written through $m_{H^{\pm }}$
and $\Delta S_{1}^{2}$ variables, we can verify the compatibility level
under electroweak observables of regions once studied from vacuum behavior.
Figure \ref{fig:OP} shows the oblique parameter constraints from the
electroweak precision and how it translates data into constraints on the masses or their splittings for the extended sector for $U(1)$ and $Z_{2}$ models. In these two
cases, there are regimes compatible between the experimental fits and the
inert-2HDM predictions over $S,T$ parameters, so that a variety of model configurations exhibits an intimate relation with the electroweak precision observables \cite{STIHDM}.

Splittings between $m_{A^{0}}$ and $m_{H^{0}}$ characterized by
their respective ratio $k_{S}\equiv m_{A^{0}}/m_{H^{0}}$ have a high
level of compatibility when $\Delta S_{1}^{2}\to 0$ when $k_{S}$ is close to degeneracy. For $k_{s}>1$, compatible zones are reduced when $k_{S}$ increases, being large splittings in $\Delta S_{0}^{2}$ compensated with large splittings in $\Delta S_{1}^{2}$, which are suppressed by perturbativity analyses. The quasi-degeneracy between neutral states is excluded for large splittings among them and charged Higgs mass. In the $U(1)$ case, 99$\%$ fit contours are approximately symmetric in $\Delta S_{1}^{2}$ splittings, implying that $ST$ parameters do not distinguish relative sign between sum of neutral states masses and charged Higgs mass. For $\Delta S_{2}^{2}$ close to zero only splittings with $\Delta S_{1}^{2}\approx0$ are allowed. Hence, in the particular limit of $m_{22}^{2}\simeq m_{H^{\pm}}^{2}$, a \emph{compressed} scenario for IHDM (quasi-degeneracy in masses of the inert scalar states) is favored by systematics of $ST$ oblique parameters.

Above all, it seems pertinent to point out that vacuum analysis as well as oblique parameters allow to determine space parameters compatible with phenomenology coming from colliders searches and dark matter studies. In the first case, in the quasi-degeneracy case of neutral states, LEP II analysis excludes the region of masses where simultaneously: $m_{H^{0}}<80$ GeV, $m_{A^{0}}<100$ GeV and $m_{A}-m_{H}>8$ GeV ($m_{H^{0}}(\text{ GeV})>8/(k_{s}-1)$, with $k_{s}=m_{A^{0}}/m_{H^{0}}$) \cite{LEPII}.  For $m_{H^{0}}( \text{ GeV})<8/(k_{s}-1)$, the LEP I limit $m_{H^{0}}+m_{A^{0}}>m_{Z^{0}}$ applies, preventing invisible $Z^{0}\to A^{0}H^{0}$ channel \cite{143Sher, 147Sher}. In terms of $h^{0}$ mass and for $m_{A^{0},H^{0}}<m_{h^{0}}$, precision tests predict that $h^{0}\to A^{0}A^{0}$ and $h^{0}\to H^{0}H^{0}$ decays shall be dominant channels. In particular, for the $U(1)$ case both decays are invisible, meanwhile in the $Z_{2}$ case the $h^{0}\to H^{0}H^{0}$ decay will be the invisible one. These effects can be ruled out by the Run 2 in the LHC, if the properties of the scalar Higgs boson with $m_{h}=125$ GeV  are still compatible with the ones predicted by SM.

For the charged Higgs boson and owing to the kinetic-gauge interactions, the dominant decays are $H^{\pm}\to W^{\pm}H^{0}$ and $H^{\pm}\to W^{\pm}A^{0}$. In degeneracy limit, both channels can be distinguished by precision tests over parity and spin of possible subsequent final decays \cite{CMSATLAS}. In the $Z_{2}$ case, $m_{A^{0}}>m_{H^{\pm}}$ hierarchy forbids the last decay channel at least as an on shell one. Nevertheless, there are also trilinear gauge couplings among charged Higgs boson and neutral gauge bosons leading to new decays channels which can compete with decays involving $Z_{2}$ odd scalar states.

Finally, we discuss some consequence of our results in front of dark matter phenomenology for the IHDM. Measured relic abundance density for dark matter is $\Omega_{cdm}h^{2}=0.1199\pm0.0022$ \cite{Planck}. This selects distinct zones in the parameter space for new physics \cite{Phase}\footnote{In \cite{Phase} is discussed the possible origin of a strong phase transition in the inert-2HDM required in baryogenesis processes; being it possible when the resonant scenario is considered.}: i) \emph{Low mass regime}:  $m_{H^{0}}<m_{h^{0}}/2$. The Dark matter pair-annihilation predominantly proceeds via the pair production of the $b$ quarks and $\tau$ leptons; being SM-like Higgs boson the dark matter portal. ii) \emph{Resonant regime}: $m_{H^{0}}\sim m_{h^{0}}/2$. This scenario produces a viable mass around a pole leaving even an unconstrained window close to 10-15 GeV around of $m_{h^{0}}/2$ \cite{Phase2}.  iii) \emph{Intermediate mass regime}: $m_{h^{0}}/2<<m_{H^{0}}(\text{GeV})<500$. Here $H^{0}$ pair annihilation to gauge bosons becomes significant, such that the thermal relic density is systematically
below the universal dark matter density for any combination of model parameters, excluding the presence of dark matter constituents \cite{13Phase,21Phase}. Moreover, as the charged Higgs holds $m_{H^{\pm}}>m_{H^{0}}$ in both models (with $Z_{2}$ and $U(1)$ symmetries), zone approaching to the upper bound is also incompatible with EW-ST parameters. iv) \emph{Heavy mass regime}: $m_{H^{0}}\gtrsim 500$ GeV. This scenario, in the lower bound, can be rendered as a decoupling among $Z_{2}$ odd scalars and $h^{0}$ wherein there is no relic density enough. If couplings with $h^{0}$ are driven out away from zero cancellations, it is possible leads to achieve the correct relic density for mass values from the lower bound slightly different up to heavy scalar settled in TeV scale.  Again incompatibility of this scenario comes from oblique quantities and unitarity constraints, which can be evaded if this inert model is considered as an effective theory of a strong interacting sector with new physics set up in the TeV energy \cite{DarkPhase}. Taking into account the $U(1)$ case, all these constraints in the different scenarios must also be satisfied by pseudoscalar Higgs boson. This fact can yields discrepancies in the matching of the relic density value since direct dark matter with spin-independent searches put limits over the degeneracy between $H^{0}$ and $A^{0}$ \cite{Kopp, 146Sher}. Nevertheless, from quantum gravity analyses, a recent approach \cite{Farinaldo} has discarded likely dark matter candidates for a 2HDM with a global $U(1)$-symmetry.

\section{Remarks and Conclusions}

Benchmarks scenarios for physics beyond the SM have changed from the
Higgs boson discovery by CMS and Atlas collaborations in LHC. The region
mass compatible with the scalar signal measured is around $125$
GeV. This scale shows outstanding features related to vacuum behavior
according to the model background. In minimal SM, computations at NNLO
exclude absolute stability at 95$\%$ C.L. for the current mass region in $m_{t}-m_{h};$ plane, showing a preference for a metastable Higgs potential for high
energy scales. Hence the Higgs self-coupling approaches to zero in Planck
energy ranges, implying a critical phase that could be explained by
either dynamical or symmetry reasons. The former argument is related to new fields
interaction at vanishing scale even as threshold corrections, meanwhile the
last fact arises from radiative corrections for classical Lagrangian
parameters.

In our studies, these possibilities are encoded in extended Higgs sectors,
where threshold vacuum behavior comes from corrections at one loop level for
2HDM type I with one inert doublet (i.e. $\langle \Phi _{2}\rangle _{0}=0$).
We are additionally taking into account $U\left( 1\right) $ and $Z_{2}$ global symmetries for the Higgs potential since these transformations preclude the occurrence of FCNCs processes at tree level. FCNCs mechanisms are highly constrained by
experiments like meson oscillations. The global $U(1)$ symmetry yields a degeneracy between $A^{0}$ and $H^{0}$. Even though,
degeneracy among neutral eigenstates is also avoided by the presence of $\lambda
_{5}$ coupling, which comes from considering a general $Z_{2}$ symmetric Higgs
potential. 

Since 2HDMs contain a bigger parameter space, a strong first order phase transition could take place, which is relevant to achieve a successful BAU via baryogenesis. Thus, they are important models to address the matter-antimatter asymmetry of the universe. This remark is an additional motivation to study vacuum structures of 2HDMs at tree level and with radiative corrections since our analyses are inspired in quantifying the threshold effects for stability
and to explore their impact on the Higgs sector for these two limiting models.

The constraints at tree level for a bounded from below potential are
considered with the traditional inequalities, which evaluate different unstable or stable zones in all directions of the field space. In addition to the standard constraints over quartic couplings of
the Higgs potential, in the $Z_{2}$-symmetry case we analyze $\lambda _{4}+\lambda
_{5}<0$ and $\lambda_{4}-\lambda_{5}>0$ inequalities. The last conditions are
translated into masses through $m_{A^{0}}>m_{H^{\pm }}>m_{H^{0}}$
restriction. On the other hand, for the Higgs potential with a $U\left(
1\right) $ symmetry, this condition for scalar spectrum implies $\lambda _{4}<0$; bringing to the charged Higgs to be the heaviest scalar state of the inert-2HDM since in this scenario exist a degeneracy among neutral states $m_{H^{0}}=m_{A^{0}}$.

For the $U\left( 1\right) $ and $Z_{2}$ cases, contours in the planes $%
\lambda _{1}\left( m_{Z}\right) -\lambda _{3,4,5}\left( m_{Z}\right) $ were considered when the vacuum positivity relations are elevated to be accomplished with the effective quartic couplings in the Higgs potential at one loop level. Those structures allow studying the new sources of instabilities in the $\Phi _{1},\Phi _{2}$ directions or in the $\Phi _{1}-\Phi _{2}$ plane of the 2HDM-field space. Finally, they are translated into constraints over scalar masses, or more particular, in splitting between them, where the
discovered scalar state in LHC has been identified with the lightest scalar CP even of the inert-2HDM. The last regime is interpreted as the alignment limit, where the mass scale of the remaining scalars could even be at EW scale. Fixing the minimality principle and vacuum behavior of the model, we look for the scalar mass values
compatible with perturbative unitarity constraints and EW precision tests by
oblique EW-parameters realization. 

The 2HDM-type I threshold corrections at one loop increase energy scale for Higgs potential stability by the introduction of new fields and couplings among them, all compared with the vacuum behavior in the SM minimal (with Higgs masses of the order of the central value of the current experimental
signal $m_{h}=125.04$ GeV). However, new sources of instabilities in those scales appear in the $\Phi _{1}-\Phi _{2}$ plane by the evolution of the remaining quartic couplings, which has many implications for the behavior of mass eigenstates.
For instance, in the splitting between states $m_{H^{\pm }}^{2}$ and $\bar{m}_{22}^{2}$ evolution
(encoded in $\lambda _{3}$), is shown as positive zones are favored for
the reference value in EW scale ($m_{Z}$). Moreover, this zone is also compatible with
perturbative unitarity behavior of scalar scattering. All results favor the scenario in which the charged Higgs is the heaviest mass state present in the $U(1)$ invariant
Higgs potential for an inert vacuum. Meanwhile, $A^{0}$ is the heaviest one in $Z_{2}$ theory, with $m_{A^{0}}>m_{H^{\pm}}>m_{H^{0}}$. Both statements come mainly from also avoiding a charge violation minimum.

Behavior of SM parameters could be extrapolated to the inert 2HDM. Particularly,
in the 2HDM type I, strong instability sources come from $\lambda
_{1}\left( \mu \right) $ evolution. These instability zones could be present
even in $\mu=10^{3}$ GeV for some zones of the parameter space. For
instance, in $\Delta S^{2}_{1}$ (when $\lambda _{1}\left( m_{Z}\right)
=m_{h^{0}}^{2}/v^{2}\simeq 0.258)$, stability zone is located at $-0.06\lesssim
\lambda_{4}(m_{Z})\lesssim 0$. However, these zones near to $\Delta S^{2}_{1}=0$ could be evaluated with custodial symmetry behavior at one loop level, being
this fact determined from $ST$-electroweak parameters (for $U=0$).
Meanwhile, non-critical values for $\lambda _{1}\left( \mu \right) $ at
Planck scales are compatible with $S-T$ fit contours at 99$\%$ C.L. in values $m_{H^{\pm}}^{2}>m_{22}^{2}$. In the $Z_{2}$ scenario, analyses favored  small splittings between the mass of $A^{0}$ and $H^{0}$, making more compatible the $m_{A^{0}}^{2}/m_{H^{0}}^{2}>1$ condition. The compressed scenario, with an approximated degeneracy between $m_{A^{0}}$ and $m_{H^{0}}$ is not ruled out in the $S-T$ plane at $99\%$ C.L. for $\Delta S_{1}^{2}<0.5$ and $\Delta S_{0}^{2}<0.25$. Nonetheless, presence of a global minimum analyses exclude this compressed regime for $\vert m_{22}^{2}\vert>2500$ GeV$^{2}$. Finally, EW-global minimum belongs in $m_{H^{\pm}}^{2}>m_{22}^{2}$, which is in consistency with vacuum analyses.

Vacuum stability systematics discriminate between the most general form for Higgs potentials in both cases, with $Z_{2}$ and $U(1)$ symmetries, since hierarchy for mass eigenstates obtained is distinct from those models. From the above discussion, in the first case the pseudoscalar $A^{0}$ arises as the heaviest scalar particle, meanwhile in the $U(1)$ model, charged Higgs boson plays this role. Hence in the $Z_{2}$ case, the most natural dark matter candidate is $%
H^{0}$ and, by contrast, in the $U(1)$ case both $A^{0}$ and $H^{0}$ might be good prospects. However, models with a $U(1)$-symmetry have been ruled out from quantum gravity analyses \cite{Farinaldo}, excluding the presence of likely dark matter candidates for these models with abelian global symmetries.

In the inert-2HDM happens that the parameter space compatible with the simultaneous existence of both vacua is larger than the predicted by tree-level studies \cite{FerreiraMet}. In this direction, and using reparametrization group theorems, we describe new discriminants to find one EW-global minimum at tree-level. It can be a useful aid to investigate the metastable behavior at NLO since at this order is possible that the nature of vacuum change at one loop level concerning to the established at tree-level. The last can be interpreted in the sense of how quantum corrections trigger phase transitions between Inert and Inert-like vacuum structures. Hence possible zones investigated by vacuum stability and precision observables can be excluded by the presence of an inert-like vacuum at one loop level. This fact is an important issue that should be addressed using properties here computed about features of a global minimum at tree level.

Within our simple framework, these limits provide a test into the scale characterizing possible sources of instabilities or new physics appearance in several zones of the parameter space in both cases. This systematic lead to determine the evolution of unstable scales for different regions of field space, complementing previous studies in vacuum behavior in the Inert 2HDM. However, these studies deal open questions about the possible
additional threshold contributions from the 2HDM to explain the boundaries separating metastability and stability in SM (the so-called ``criticality'') from the input of more general Higgs potentials. Nevertheless, when more precision tests become
performed at LHC, and with most accurate values of parameters, an extension to analyses
must be introduced to explain issues related to the behavior of effective Higgs potential, baryonic asymmetry of the universe, and the dark matter origin. Perhaps in
those scenarios, higher radiative corrections beyond NLO for 2HDM couplings
should be considered and hence studies about vacuum nature and its behavior might be completed.

\section*{Acknowledgments}

We acknowledge financial support from Colciencias and DIB (Universidad Nacional de Colombia). Particularly, Andr\'es Castillo, John Morales and Rodolfo Diaz are indebted to the \emph{Programa Nacional Doctoral of Colciencias} for its academic and financial support. Andr\'es Castillo also thanks kind hospitality of the Theoretical Physics Department and the Group of Effective Theories in Modern Physics at the Universidad Complutense de Madrid (Spain), where some results of this paper were written and broadly discussed. Carlos G. Tarazona also thanks to the financial aid from DIB-Project with Q-number \textbf{110165843163} (Universidad Nacional de Colombia).

\appendix

\section{Renormalization Group Equations for $Z_{2}-$2HDM \label{ap:RGEsection}}

The behavior of the parameters (couplings) and relations among
them are computed through the Renormalization Group Equations (RGEs). At
higher levels in perturbation theory, quartic couplings depend
on the energy scale $\mu $. To evaluate the presence of
instabilities in all field space,  we demand that energy scale-dependent couplings
satisfy the same constraints obtained at tree level in section \ref%
{sec:TL}, ensuring an effective Higgs potential bounded from below and, hence, preventing a possible decay of EW minima.

Besides of the importance for the vacuum behavior analyses, the RGEs are valuable tools to determine (by the triviality principle) energy bounds of the parameters and perturbative validity of the model. To evaluate the $\lambda_{i}(\mu)$ at one loop level,  the RGEs of all remaining couplings must be considered simultaneously. The remaining couplings correspond to the gauge group couplings $g^{\prime },g,$ $g_{s}$ of the symmetry groups $U(1)$, $SU(2)$, $SU(3)$ and the third generation of Yukawa couplings $\eta _{tt}$ (top), $\eta _{bb}$ (bottom) and $\eta _{\tau \tau }$. All Yukawa's elements are coupled to the $\Phi _{1}$ doublet uniquely. All of them are computed in Refs \cite{RGE,RGE01}. The one
loop RGEs for a general gauge theory are also presented in \cite{RGE1}-\cite{RGE1b} and for
NHDM (N Higgs Doublet Model) with a gauge group $SU(2)_{L}\times U(1)_{Y}$, and they were computed in \cite{RGE2,RGE3}. For the particular inert-2HDM, 
RGEs can be found in \cite{Phase}, which have been proved with \texttt{SARAH}-package \cite{SARAH} and \texttt{Pyr@te}-package \cite{Pyrate}.

We summarize the RGEs for 2HDM with SM gauge group through the respective
settlement of equations for $Z_{2}$ global invariant Higgs
potential. The Renormalization Group Equations for the gauge sector at one loop
level are given by

\begin{eqnarray}
\frac{dg}{dt} &=&\frac{1}{16\pi ^{2}}\left( \frac{4}{3}n_{f}+\frac{1}{6}%
n_{H}-\frac{22}{3}\right) g^{3}=-3g^{3}, \\
\frac{dg^{\prime }}{dt} &=&\frac{1}{16\pi ^{2}}\left( \frac{20}{9}n_{F}+%
\frac{1}{6}n_{H}\right) g^{^{\prime }3}=7g^{^{\prime }3}, \\
\frac{dg_{s}}{dt} &=&\frac{1}{16\pi ^{2}}\left( \frac{4}{3}n_{f}-11\right)
g_{s}^{3}=-7g_{s}^{3}.
\end{eqnarray}%
For the 2HDM, $n_{H}=2$ and $n_{f}=3$ (the same fermionic content of SM)$.$
In all equations $t=\log \mu .$ For the 2HDM type I, we can summarize the RGEs
for Yukawa couplings in the heaviest fermions ($\tau ,b,t$) by

\label{YI} 
\begin{align}
16\pi ^{2}\frac{d\eta _{\tau \tau }}{dt}& =-\left( \frac{9}{4}g^{2}+\frac{15%
}{4}g^{^{\prime }2}\right) \eta _{\tau \tau }+T_{11}\eta _{\tau \tau }+\frac{%
3}{2}\eta _{\tau \tau }^{3}, \\
16\pi ^{2}\frac{d\eta _{bb}}{dt}& =-\left( 8g_{s}^{2}+\frac{9}{4}g^{2}+\frac{%
5}{12}g^{^{\prime }2}\right) \eta _{bb}+T_{11}\eta _{bb}+\frac{3}{2}\eta
_{bb}^{3}-\frac{3}{2}\eta _{tt}^{2}\eta _{bb}, \\
16\pi ^{2}\frac{d\eta _{tt}}{dt}& =-\left( 8g_{s}^{2}+\frac{9}{4}g^{2}+\frac{%
17}{12}g^{^{\prime }2}\right) \eta _{tt}+T_{11}\eta _{tt}+\frac{3}{2}\eta
_{tt}^{3}-\frac{3}{2}\eta _{bb}^{2}\eta _{tt}.
\end{align}%
Since the Yukawa matrices are diagonal, $T_{11}=3\left( \eta _{tt}^{2}+\eta
_{bb}^{2}\right) +\eta _{\tau \tau }^{2}.$ Here $\Phi _{2\text{ }}$has been
decoupled from all fermions. Here top quark, bottom quark and $\tau $ lepton
are only coupled to the $\Phi _{1}$ doublet. The RGE for scalar couplings ($%
Z_{2}$ global invariant Higgs potential) at one loop in a 2HDM type I are
the following set of differential equations

\begin{align}
16\pi ^{2}\frac{d\lambda _{1}}{dt}& =12\lambda _{1}^{2}+4\lambda
_{3}^{2}+4\lambda _{3}\lambda _{4}+2\lambda _{4}^{2}+2\lambda _{5}^{2}+\frac{%
9g^{4}+6g^{2}g^{^{\prime }2}+3g^{^{\prime }4}}{4}  \notag \\
&-\left( 9g^{2}+3g^{^{\prime }2}\right) \lambda _{1} -6\left( \eta
_{bb}^{4}+\eta _{tt}^{4}\right) -2\eta _{\tau \tau }^{4}+4\left( 3\left(
\eta _{tt}^{2}+\eta _{bb}^{2}\right) +\eta _{\tau \tau }^{2}\right) \lambda
_{1}, \\
16\pi ^{2}\frac{d\lambda _{2}}{dt}& =12\lambda _{2}^{2}+4\lambda
_{3}^{2}+4\lambda _{3}\lambda _{4}+2\lambda _{4}^{2}+2\lambda _{5}^{2}+ 
\frac{9g^{4}+6g^{2}g^{^{\prime }2}+3g^{^{\prime }4}}{4}-\left(
9g^{2}+3g^{^{\prime }2}\right) \lambda _{2}, \\
16\pi ^{2}\frac{d\lambda _{3}}{dt}& =2\left( \lambda _{1}+\lambda
_{2}\right) \left( 3\lambda _{3}+\lambda _{4}\right) +4\lambda
_{3}^{2}+2\lambda _{4}^{2}+2\lambda _{5}^{2}+ \frac{9g^{4}-6g^{2}g^{^{\prime
}2}+3g^{^{\prime }4}}{4}  \notag \\
&-\left(9g^{2}+3g^{^{\prime }2}\right) \lambda _{3}+2\left( 3\left( \eta
_{tt}^{2}+\eta _{bb}^{2}\right) +\eta _{\tau \tau }^{2}\right) \lambda _{3},
\\
16\pi ^{2}\frac{d\lambda _{4}}{dt}& =2\lambda _{4}\left( \lambda
_{1}+\lambda _{2}\right) +2\left( 2\lambda _{4}^{2}+4\lambda _{3}\lambda
_{4}\right) +8\lambda _{5}^{2}+3g^{2}g^{^{\prime }2} -\left(
9g^{2}+3g^{^{\prime }2}\right) \lambda _{4}\notag\\
&+2\left( 3\left( \eta
_{tt}^{2}+\eta _{bb}^{2}\right) +\eta _{\tau \tau }^{2}\right) \lambda _{4},
\\
16\pi ^{2}\frac{d\lambda _{5}}{dt}& =2(\lambda _{1}+\lambda _{2})\lambda
_{5}+8\lambda _{3}\lambda _{5}+12\lambda _{4}\lambda _{5}-\left(
9g^{2}+3g^{^{\prime }2}\right) \lambda _{5}+2\left( 3\left( \eta
_{tt}^{2}+\eta _{bb}^{2}\right) +\eta _{\tau \tau }^{2}\right) \lambda _{5}
\label{RGEI}
\end{align}%

Because of fermions are coupled to one and only one doublet, we can expect several
contributions to instabilities from $\lambda _{1}$ coupling associated with the
quartic coupling of $\Phi _{1}$ dimension four operators. Also, high values of all couplings in their initial conditions evaluated in $\mu=m_{Z}$ might drive out to sources for non-perturbative scenarios. These initial conditions over parameter space will be related to the central values for $m_{h^{0}},m_{t},m_{b},m_{W}$ and $m_{Z}$ given in \cite{PDG}.

\end{document}